\documentclass{aa} 

\usepackage{amsmath}
\usepackage{graphicx}
\usepackage{subcaption}
\usepackage{mwe}
\usepackage{graphicx}
\usepackage{txfonts}

\usepackage{xcolor}
\usepackage{comment}
\usepackage{url}

\usepackage{hyperref}  
\begin{document} 
\title{Wavelet-based statistics for enhanced 21cm EoR parameter constraints}

\titlerunning{Wavelet-based statistics for parameter constraints}

  \author{Ian Hothi 
     \inst{1,}\inst{2}\fnmsep\thanks{Ian.Hothi@obspm.fr}
     \and
     Erwan Allys\inst{2}
     \and
     Benoit Semelin\inst{1}
     \and
     Francois Boulanger\inst{2}
     }

  \institute{LERMA, Observatoire de Paris, PSL Research University, CNRS, Sorbonne Universit\'{e}, F-75014 Paris, France
     \and
       Laboratoire de Physique de l’ENS, ENS, Universit\'{e} PSL, CNRS, Sorbonne Universit\'{e}, Universit\'{e}e Paris Cit\'{e}, 75005 Paris, France
       }

  \date{Received 31 October 2023 / Accepted 11 March 2024}

 
\abstract{
We propose a new approach to improve the precision of astrophysical parameter constraints for the 21cm signal from the epoch of reionisation (EoR). Our method introduces new sets of summary statistics, hereafter `evolution compressed' statistics, which quantify the spectral evolution of the 2D spatial statistics computed at fixed redshift. We defined such compressed statistics for power spectrum (PS), wavelet scattering transforms (WST), and wavelet moments (WM), which also characterise non-Gaussian features. 
To compare these different statistics, along with the 3D power spectrum, we estimated their Fisher information on three cosmological parameters from an ensemble of simulations of 21cm EoR data, both in noiseless and noisy scenarios using Square Kilometre Array (SKA) noise levels equivalent to 100 and 1000 hours of observations. We also compare wavelet statistics, in particular WST, built from standard directional Morlet wavelets, as well as from a set of isotropic wavelets derived from the binning window function of the 2D power spectrum.
For the noiseless case, the compressed wavelet statistics give constraints that are up to five times more precise than those obtained from the 3D isotropic power spectrum. At the same time, for 100h SKA noise, from which it is difficult to extract non-Gaussian features, compressed wavelet statistics still give over 30\% tighter constraints. 
We find that the wavelet statistics with wavelets derived from the power-spectrum binning window function provide the tightest constraints of all the statistics, with the WSTs seemingly performing better than the WMs, in particular when working with noisy data. 
The findings of this study demonstrate that evolution-compressed statistics extract more information than usual 3D isotropic power-spectra approaches and that our wavelet-based statistics can consistently outmatch power-spectrum-based statistics. 
When constructing such wavelet-based statistics, we also emphasise the need to choose a set of wavelets with an appropriate spectral resolution concerning the astrophysical process studied.}


\maketitle

%

\section{Introduction}
The formation of the first stars marked a pivotal phase change in the Universe, known as the epoch of reionisation (EoR), during which the ionisation state transitioned from a predominantly neutral state to a primarily ionised one. This significant epoch serves as the precursor to the state of the intergalactic medium (IGM) that we observe today. While direct observations of the EoR are challenging, indirect constraints on its timings and progression have been obtained.
Observations of the cosmic microwave background (CMB) have provided valuable insights, and suggest that a substantial portion of the reionisation process occurred after a redshift of $z = 14$ \citep{PlanckEoR16, 2018:GorceDouspisAghanim}. Furthermore, the prominence of the Gunn-Peterson trough observed in the spectra of quasars has offered additional constraints. Early investigations by \citet{GPTroughFan} indicated that a considerable fraction of the Universe underwent ionisation by a redshift of $z\sim 6$. However, more recent findings based on a larger number of spectra have lead to an evolution of this perspective, suggesting a later completion of reionisation around $z\sim 5.3$ \citep{2015:BeckerBoltonMadau, 2018:BosmanFanJiang, 2021:QinMesingerBosman}. 

The 21cm signal emitted from the hyperfine transition in neutral hydrogen during the EoR is the most promising observable to constrain the EoR. The most common statistic used to constrain the EoR's 21cm signal is its 3D spherical power spectrum. Several radio telescopes that are either currently taking data or are yet to come online are designed to detect the 21cm power spectrum; for example, the
Low-Frequency Array (LOFAR)\footnote{http://www.lofar.org/} \citep{LOFAROVERVIEW}; the Giant Metrewave Radio Telescope (GMRT)\footnote{http://gmrt.ncra.tifr.res.in/} \citep{GMRT1,GMRT2}; the Murchison Widefield Array (MWA)\footnote{http://www.mwatelescope.org/} \citep{MWAPaper}; the Hydrogen Epoch of Reionization Array (HERA)\footnote{http://reionization.org} \citep{HERA}; the new extension in Nan\c{c}ay upgrading LOFAR (NeNuFar)\footnote{https://nenufar.obs-nancay.fr/en/homepage-en/} \citep{NenuFAR,2023:Munshi}; and the Amsterdam-ASTRON Radio Transients Facility and Analysis Center (AARTFAAC)\footnote{http://aartfaac.org} \citep{AARFAAC_16,gehlot2019}.

The power spectrum, which quantifies signal power distribution as a function of scale, serves as a valuable two-point statistic for analysing the 21cm signal from the EoR. To probe the EoRs non-Gaussian nature, which is related to coupling between scales, is not possible solely with the power spectrum. In this context, the three-point statistics, such as the bispectrum, has garnered significant attention within the EoR field \citep{Shimabukuro2016a, Majumdar2017, Watkinson2017, 2019:GorcePritchard, 2020:HutterWatkinsonSeiler, 2021:WatkinsonTrottHothi}. These high-order statistics are employed to capture more of the complexities of the signal. The bispectrum, derived from the Fourier transform of three-point correlations, characterises the statistical dependence among triplets of distinct sets of Fourier modes in a signal. Consequently, it allows us to go beyond power spectrum studies.
As a high-order statistic, the bispectrum encounters challenges, such as high levels of variance due, in particular, to the presence of outliers \citep{2005:Welling}. Outliers are extreme values that significantly deviate from the typical values of the data and can have a disproportionate impact on high-order moments. The bispectrum is sensitive to outliers because it is a high-order computation that involves triple products. Such high levels of variance in our summary statistics will make it more difficult to precisely infer physical parameters from limited data.
To overcome these limitations, novel wavelet-based statistics based on wavelet transforms have emerged \citep{2011:Mallat}. Wavelet transforms have been used in cosmological parameter estimation in the form of wavelet moments (WM), which apply statistical moments to extract information from wavelet transform; see for example \citet{2022:EickenbergAllysMoradinezhadDizgah}. More hierarchical representations ---using a cascade of wavelet transforms and non-linearities--- called wavelet scattering transforms (WST) have also been applied to cosmology and found successful applications in studying the highly non-Gaussian interstellar medium (ISM), where the interplay between gravity, magneto-hydrodynamics, and various processes within the ISM generates complex statistical patterns \citep{2019:AllysLevrierZhang, 2020:Regaldo-SaintBlancardLevrierAllys,2021:SaydjariPortilloSlepian}. WSTs have also been used for classification and parameter inference in the context of weak lensing \citep{2021:ChengMenard} and the large-scale structure of the Universe \citep{2022:ValogiannisDvorkin}. These scattering transforms offer a means to probe non-Gaussianity without suffering from the shortcomings associated with high-order statistics, such as the bispectrum.
\citet{2022:GreigTingKaurov} were the first authors to apply WST to 2D images of the 21cm signal across multiple redshifts. Using a Fisher formalism for astrophysical parameter constraints, these authors demonstrated that the WST performs comparably to, and in some instances better than, a 3D power spectrum analysis on the same dataset. This latter result is a pioneering study in its application of the WST to the 21 cm signal and as such should be confirmed and extended.

In their comparison of the power spectrum and the WST, 
\citet{2022:GreigTingKaurov} found that certain aspects appear to favour the power spectrum, potentially skewing the results. Specifically, the authors employed a Fourier bin range for the power spectrum that effectively mitigates resolution effects within the simulations. Furthermore, they used a binning scheme that particularly benefits larger scales, which are less affected by noise. However, a
comparable treatment is not extended to the WST, which may be more impacted by noise.

Another point is that the calculation of the covariance for the 21cm power spectrum, which included 1000 hours of SKA noise, involved the use of {\fontfamily{cmtt}\selectfont
{21CMSENSE}}\footnote{\url{//github.com/jpober/21cmSense}}\citep{2013:PoberParsonsDeBoer,2014:PoberLiuDillon} and thus (i) does not contain off-diagonal contributions, from cosmic variance for example (see \citet{2023:PrelogovicMesinger}), and (ii) is sensitive to the limited number of simulations used in the study. In contrast, the covariance for WST was directly derived from the limited number of fiducial simulations. This discrepancy in the covariance estimation methods introduces a potential bias.

In the present paper, we introduce 2+1 statistics ---hereafter referred to as {evolution-compressed} statistics---, which quantify the spectral evolution of the 2D spatial statistics computed at a fixed redshift in order to enhance the constraint on astrophysical parameters. With these evolution-compressed summary statistics, we aim to extract additional information beyond the 3D isotropic power spectrum, which is a well-established and extensively studied statistic in 21cm physics. We achieved this using recently developed wavelet-based statistics, which we built from a set of wavelets with a proper spatial frequency sampling. We validated this set of wavelets by first recovering the same information as the power spectrum in a wavelet framework, before adding non-Gaussian information through the use of non-linear statistics. We compared these wavelet statistics, built both from the WST statistic and WM, to the 2D evolution compressed power spectrum, and the 3D spherically averaged power spectrum; this comparative analysis provides valuable insights into the potential benefits and limitations of incorporating anisotropic summary statistics into parameter estimation.

The paper is structured as follows: We introduce the statistics used in this work ---including WST and how its reduced form, the reduced wavelet scattering transform (RWST), is calculated--- and our wavelet application in Section \ref{Stats}. Section \ref{Simulation_Set_Up} outlines the simulation used in this paper and how we set up our Fisher analysis. In Section \ref{Results}, we show the results of comparing our different summary statistics under different noise treatments. We then summarise our findings and provide conclusions
in Section \ref{Conclusions}.

\section{Statistics}
\label{Stats}

In this section, we present the various summary statistics used in this paper, including the 2D and 3D power spectrum, WM, and WST. Except for the 3D power spectrum, these statistics are calculated for fixed-$z$ 2D spatial slices of the lightcone. Towards the end of this section, we also introduce the compression method that we use to compress the redshift evolution for 2D spatial statistics.

\subsection{Power spectrum}
\label{Power_Spectrum}

The power spectrum is the most commonly used statistic in the 21cm field. We consider a field $I(\vec{x})$, where $\vec{x}$ refers to the spatial location, as well as $\Tilde{I}(\vec{k})$ its Fourier transform. The estimated power spectrum for a given bin is:
\begin{equation}
\label{eq:ps}
  P_i = \int | \Tilde{I}(\vec{k})\cdot W_i (\vec{k})|^2 d\vec{k},
\end{equation}
where $W_i (\vec{k})$ is the spectral window function defining the $i^{th}$ bin. The power spectrum is then an ensemble of $N_b$ summary statistics, with $N_b$ being the number of bins. In this paper, we restrict ourselves to isotropic power-spectrum estimates, which correspond to spectral windows for which $W_i(\vec{k})$ depends only on $k$ the modulus of $\vec{k}$. A typical choice for this window function is the top hat function, which equally weighs all $k$-modes in a given bin. 

In this paper, we explore two applications of the power spectrum. 
The first application involves a first step of 2D spectral binning of the lightcone, that is, applying the power spectrum to each given frequency channel. As we consider an isotropic power spectrum, the window function, $W_i (\vec{k})$, is only a function of ${k} = \sqrt{\vec{k}^2_x+\vec{k}^2_x}$, and corresponds to 2D circular shells. We chose to use a Gaussian window function (see Section \ref{Wavelet_Moments} for the reasoning behind this
choice), where the spectral window function of the $i^{th}$-bin is defined as:
\begin{equation}
\label{EqSpectralWindows}
  W_i(\vec{k}) = \exp\left(-\frac{(k - \bar{k}_i)^2}{2{\sigma_i}^2}\right),
\end{equation}
where $\bar{k}_i$ is the central frequency of the $i^{th} k$-bin and $\sigma_i$ its standard deviation. We bin our Fourier space into $N_b$ = 9 logarithmic (base 10) bins for our binning scheme. 

\begin{figure}
\centering
\includegraphics[width=1\linewidth]{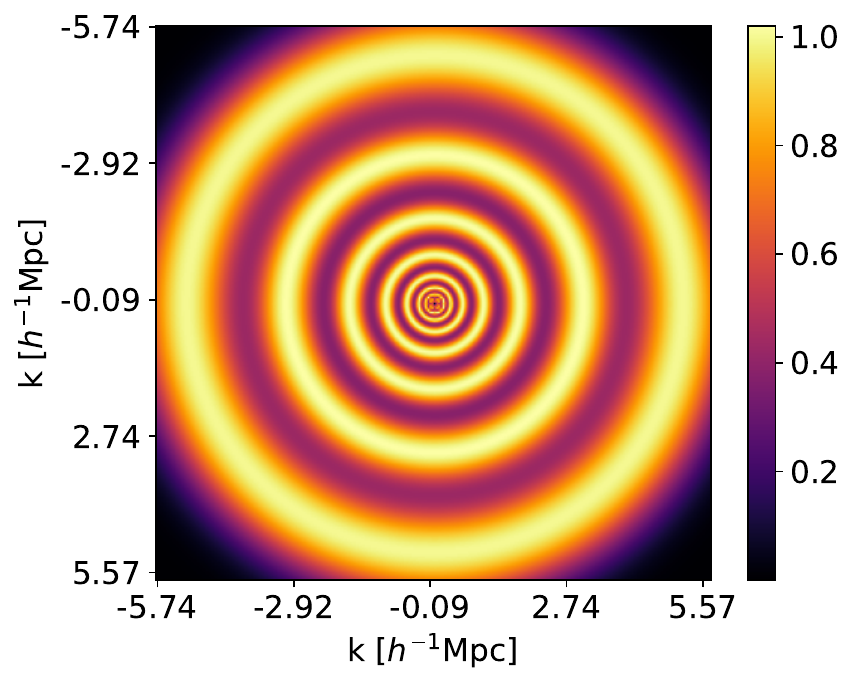}
\caption{Two-dimensional perspective of the binning window functions in Fourier space. We consider logarithmic binning, with the concentric circles close to the centre representing the window function of low $k$-bins and the concentric circles towards the edges representing the window function of high $k$-bins.}
\label{fig:WF_2D_Show}
\end{figure}

\begin{figure}
\centering
\includegraphics[width=1\linewidth]{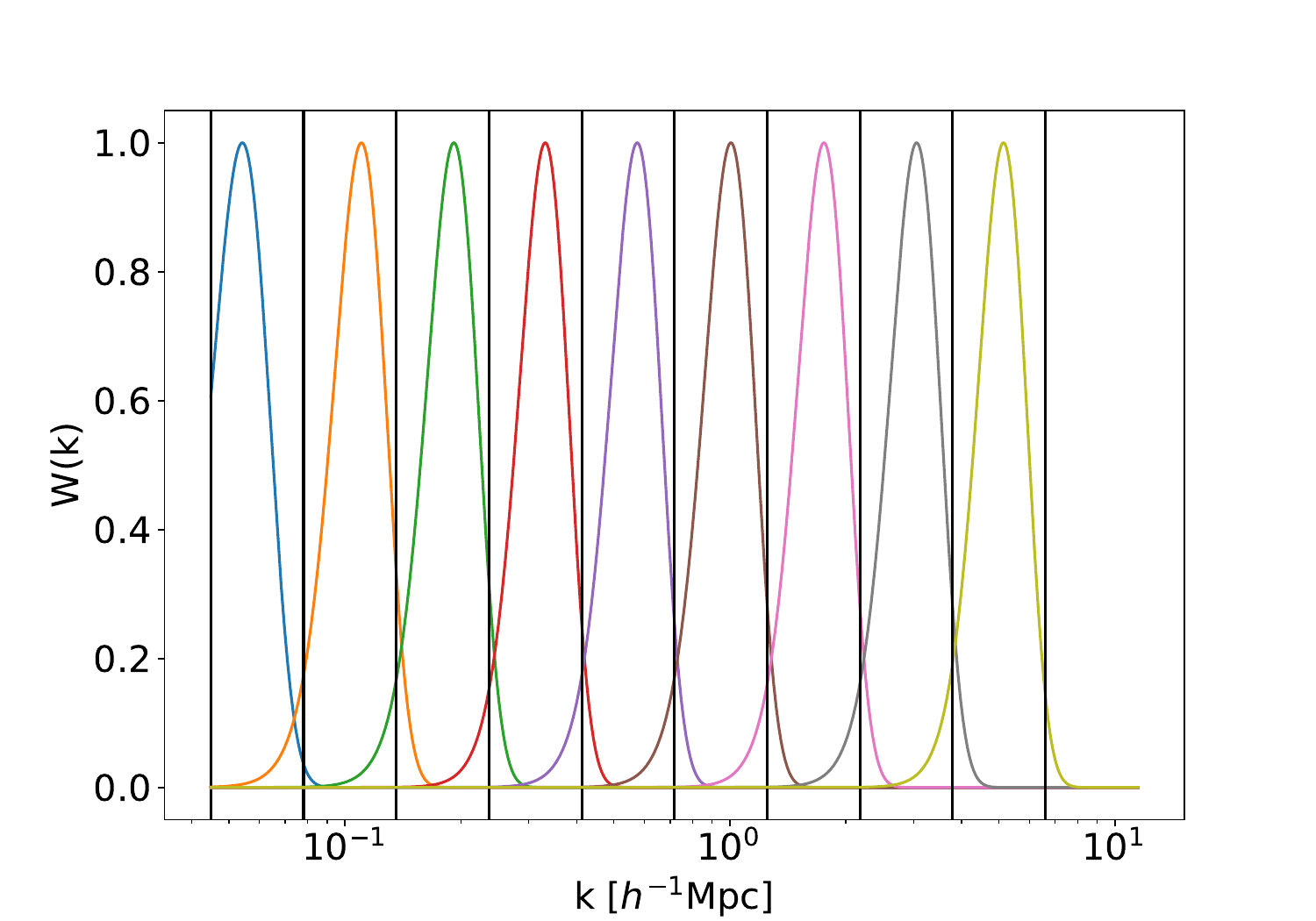}
\caption{One-dimensional line through our 2D Gaussian window functions, shown in Fig.~\ref{fig:WF_2D_Show}.}
\label{fig:WF_1D_Show}
\end{figure}

Figures \ref{fig:WF_2D_Show} and \ref{fig:WF_1D_Show} show these 2D spectral windows, as well as their 1D radial profile. We use $PS_i^{2D}$ to denote the resulting power-spectrum estimates, for which we also have $N_b =9$ coefficients. While this choice of spectra window leads to some power being distributed across adjacent bins, both sampling methods produce similar power spectra, with minor differences arising from the use of the window function. These differences are more prominent on low-$k$ bins than on higher $k$-bins, and lead to a change of up to 20$\%$ in the power in a given bin.

The second application is the 3D isotropic spherically averaged power spectrum, called $P_i^{3D}$, which involves a 3D binning of the Fourier transformed lightcone. In this case, the window function, $W_i (\vec{k})$, is also a function of ${k} = \sqrt{\vec{k}^2_x+\vec{k}^2_y+\vec{k}^2_z}$ only, and corresponds to 3D spherical shells. For continuity in the treatment of our binning, we chose to use the Gaussian window function defined in equation~\eqref{EqSpectralWindows} and bin our 3D Fourier space into $N_b$ = 9 logarithmic (base 10) bins.


\subsection{Wavelet statistics}
\subsubsection{Wavelet transform and choice of wavelets}

The wavelet statistics we use are built from wavelet transforms. Wavelet transforms are a mathematical tool used to process signals and images. They provide space and scale localisation of features, which makes them useful for compression, denoising, feature extraction, and pattern recognition. Unlike traditional Fourier-based techniques, wavelet transforms decompose data into maps that describe the contribution of different scales and positions within the data, allowing a localised representation at different resolutions. 

A wavelet is an oscillatory function with zero mean that is localised around the origin. In two dimensions, a set of wavelets are built by dilating and rotating a single mother wavelet, $\psi(\vec{x})$:
\begin{equation}
  \psi_{j,\theta}(\mathbf{x}) = 2^{-j/Q} \psi \big(2^{-j/Q}r_{\theta}^{-1}\mathbf{x}\big),
\end{equation}
where $r_{\theta}^{-1}$ is the rotation operator of angle $\theta$ and $j$ is the scale parameter. We usually consider integer numbers for scales $j$ and angle $\theta$, having $J$ scales $j$ between $1$ and $J$, and dividing $\pi$ into $L$ angles. The quality factor $Q$ controls the spectral binning of the wavelet. This binning is dyadic if $Q=1$, that is, the integer scale $j$ corresponds to a scale $\sim 2^j$ in pixels. We note that isotropic wavelets are only described by a scale parameter $j$ because they are rotation invariant. In the following, we use $\lambda$ to denote the ensemble of ($j$, $\theta$) parameters used to identify each wavelet.

A wavelet transform is obtained by convoluting an image by a set of wavelets:
\begin{equation}
WT_\lambda[I](\vec{x}) = I * \psi_\lambda (\vec{x}),
\end{equation}
where $*$ denotes a convolution. Wavelet statistics are usually constructed by successive applications of wavelet transforms and non-linear operators before an overall spatial integration. However, the choice of wavelet is critical because it drives the spatial frequency resolution of the wavelet statistics. Optimal choices depend on the application. To study this choice, we considered the following two different wavelet sets.

The first set of wavelets are dyadic directional Morlet wavelets, from which the WSTs are usually built for astrophysical applications \citep{2019:AllysLevrierZhang,2020:Regaldo-SaintBlancardLevrierAllys}. These wavelets are defined as 
\begin{equation}
\psi_m(\vec{x})=\alpha\left(e^{i \vec{n} \cdot \vec{x}}-\beta\right) \cdot e^{-|\vec{x}|^2 /\left(2 \sigma^2\right)},
\end{equation}
where $\alpha$ ensures that the wavelets have a unit $\ell^2$-norm, $\beta = \exp \left(-\sigma^2 / 2\right)$ ensures a null average, and $\vec{n}$ is a unit vector defining the oscillation direction of the mother wavelet (usually the $x$ coordinate in a $(x,y)$ plane). The Gaussian window, of size $\sigma$, localises the wavelet. The set of Morlet wavelets is then built using $Q=1$ dyadic dilation and rotation, giving $\psi_{m,\lambda}$ wavelets. In this paper, we take $J_m = 6$ and $L_m=4$ and rely on the Morlet wavelets defined in the {\fontfamily{cmtt}\selectfont pywst}\footnote{\url{https://github.com/bregaldo/pywst}} package.

These Morlet wavelets have a dyadic spatial frequency sampling defined on scales of $2^j$, which could be coarse for EoR astrophysical parameter constraints. In addition, this sampling partially favours smaller scales with respect to larger ones, which could be a problem in the presence of thermal noise, which dominates at small scales. To better study the effect of the spatial frequency sampling, we also construct another set from the power spectrum sampling. An advantage of this wavelet choice is that it can also be more directly compared with power spectrum statistics as the spatial frequency sampling used by the wavelet statistics is the same.

\begin{figure*}
\centering
\includegraphics[width=\linewidth]{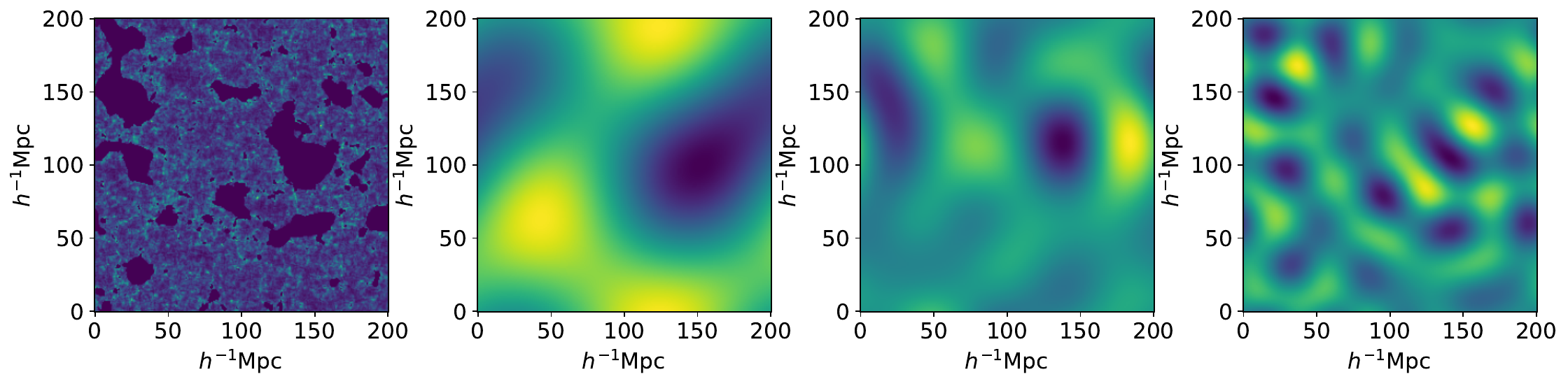}
\caption{Product of convolving a given frequency channel of a lightcone with our wavelet, ${I}(\vec{x}) * \Tilde{W}_i (\vec{x})$, which is the inverse Fourier transform of our Gaussian window function. Each wavelet is given by the inverse Fourier of a single concentric circle in Fig.~\ref{fig:WF_2D_Show}. The left-most image is of the original frequency channel with no window function applied. The rest, from left to right, are from window functions centred on: 0.113 $h^{-1}Mpc$, 0.196 $h^{-1}Mpc$, and 0.342 $h^{-1}Mpc$.}
\label{fig:Wavelet_Moment_Plots}
\end{figure*}

Usual power spectrum binning for 21cm studies uses the `top hat' spectral window function, whose spatial Fourier transform corresponds to a sinc function, with a poor spatial localisation. This explains our previous choice of a Gaussian window for the spatial frequency sampling, which is more spatially localised. As a second choice of wavelets, we thus use the inverse Fourier transform $\psi_{w,i} = \tilde{W}_i(\vec{x})$ of the previous spectral windows $W_i(\vec{k})$ defined in equation ~\eqref{EqSpectralWindows}. Contrary to the previous wavelet sets, these wavelets are isotropic and only have a scale parameter, which is called $i$ here for consistency with previous power spectrum notations. In Fig.~\ref{fig:Wavelet_Moment_Plots}, we show the application of a window function, ${I}(\vec{x}) * \psi_{w,i} (\vec{x})$, on an example frequency channel in real space. These wavelets have $J_w = 9$, for a quality factor of $Q \simeq 1.25$.

\subsubsection{Wavelet moments}
\label{Wavelet_Moments}

The first set of wavelet statistics we consider are WMs, which are integrated moments of the wavelet transforms \citep{2022:EickenbergAllysMoradinezhadDizgah}. We consider two moments, which are defined from the $\ell^1$- and $\ell^2$-norms of the wavelet convolutions\footnote{The $\ell^p$-norm $ \| x \|_p $ of a vector $\vec{x}$ is defined as 
\begin{equation}
\label{eq:lp_norm}
  \| x \|_p = \left( \sum_{i=1}^{n} |x_i|^p \right)^{\frac{1}{p}}.
\end{equation}},
yielding
\begin{equation}
  M_{1}(\lambda) = \int_{\mathbb{R}^2} | {I}(\vec{x}) * \psi_\lambda(\vec{x})| d\vec{x},
\end{equation}
and 
\begin{equation}
\label{eq:Wavelet_Moment_WF_L2}
  M_{2}(\lambda) = \int_{\mathbb{R}^2} | {I}(\vec{x}) * \psi_\lambda (\vec{x})|^2 d\vec{x}.
\end{equation}
In this paper, these moments are constructed from the $\psi_{w,i}$ wavelets, and are therefore labelled $M_{1}(i)$ and $M_{2}(i)$. 

An advantage of the $\psi_{w,i}$ choice of wavelets, which correspond to the inverse Fourier transform of the spectral window used for the power spectrum definition, is that thanks to Parseval's identity, we directly have that $M_{2}(i) = P_i$. This illustrates the direct link between the choice of wavelets when building wavelet statistics and the frequency resolution achieved. A second advantage is that, as the WM is built from the $M_1$ and $M_2$ moments, they necessarily contain more Fisher information than the $M_2$ moments, and therefore more than the power spectrum. The $M_1$ moments bring additional information about the sparsity of the field. To decorrelate the $M_1$ moments from the $M_2$ moments by as much
as possible, we normalise them as follows:
\begin{equation}
\label{eq:Wavelet_Moment_normalised_L1}
  \Bar{M_{1}}(i) = \frac{M_{1}(i)}{\sqrt{M_{2}(i)}}.
\end{equation}
The final WM statistics are the concatenation of $\Bar{M_{1}}(i)$ and $M_{2}(i)$, and thus have 2*$N_b$ coefficients.

\subsubsection{Wavelet scattering transforms}
\label{Scattering_Transforms}

The second set of wavelet statistics we consider is the WST~\cite{2011:Mallat,2019:AllysLevrierZhang}. These are constructed through a series of wavelet transforms and the application of the modulus operator, resulting in the generation of a collection of scattering coefficients. The coefficients are constructed layer by layer, and we consider only the coefficients of the first two layers, called $S_1(\lambda_1)$ and $S_2(\lambda_1,\lambda_2)$.

The first layer is constructed by convolving the 2D field $I(\textbf{x})$ with a family of wavelets $\psi_{\lambda_1}$ and applying a modulus non-linearity:
\begin{equation}
\label{eq:S1}
  S_1 (\lambda_1) = \frac{1}{\mu_1} \int |I * \psi_{\lambda_1}| (\mathbf{x}) d^2 \mathbf{x},
\end{equation}
where $\mu_1$ is a normalisation factor\footnote{ It is defined as the impulse response:
\begin{equation}
  \mu_1 = \int |\delta_{d} * \psi_{j_1,\theta_1}| (\mathbf{x})d^2\mathbf{x},
\end{equation}
where $\delta_{d}$ is the dirac delta function. The following $\mu_2$ normalisation factor is defined similarly.}. 
These coefficients are sometimes normalised by the mean of the field, but we did not do that here, because it is not suitable when the mean of the field is not a relevant quantity from a physical point of view.

The second layer is constructed by convolving the field again with another family of wavelets $\psi_{\lambda_2}$ and applying another modulus non-linearity: 
\begin{equation}
\label{eq:S2}
  S_2({\lambda_1,\lambda_2}) = \frac{1}{\mu_2} \int||I * \psi_{\lambda_1}| * \psi_{\lambda_2}(\mathbf{x})d^2\mathbf{x},
\end{equation}
where $\mu_2$ is a normalisation factor. To take into account the variability of $S_2$ due to the amplitude of the first wavelet convolution, we follow the usual normalisation by the first layer:
\begin{equation}
\label{eq:S2_norm}
  \bar{S}_2({\lambda_1,\lambda_2}) = \frac{ S_2({\lambda_1,\lambda_2})}{S_1 (\lambda_1)},
\end{equation}
where the scale $\lambda_2$ characterised by the second wavelet should be larger than the scale $\lambda_1$ characterised by the first wavelet.

In this study, we computed the WSTs from the two sets of wavelets defined above: the directional dyadic Morlet wavelets, $\psi_{m,\lambda}$, as well as those constructed from the isotropic spectral windows used for the power spectrum, $\psi_{w,i}$, and we call these WST$_m$ and WST$_w$, respectively. 

For the $WST_m$ statistics, for which coefficients are labelled by $(j,\theta)$ indices, we carry out an additional average over the direction to compute completely isotropic features. Following the results obtained with the RWST~\citep{2019:AllysLevrierZhang}, we chose to do this average on the logarithm of the coefficients. This leads us to construct \begin{equation}
S^{iso}_1=\left\langle \log_2 \bigg( S_1 (j_1,\theta_1)\bigg)\right\rangle_{\theta_1},
\end{equation}
and
\begin{equation}
\bar{S}^{iso}_2 = \left\langle\log_2 \bigg( \bar{S}_2({j_1,\theta_1,j_2,\theta_2}) \bigg)\right\rangle_{\theta_1, \theta_2}.
\end{equation}
We numerically verified that taking a logarithm before performing the angular averaging leads to better and more stable Fisher estimates. Our $S_1^{iso}$ and $\bar{S}_2^{iso,1}$ are therefore, up to $\log_2$, similar to the $S_1$ and $S_2$ definitions in \citet{2022:GreigTingKaurov}. 

For the WST$_w$, no angular averaging was necessary because they are constructed from isotropic wavelets, which do not have an angular dependency. For these second statistics, we did not take the logarithm for these statistics, because it slightly worsened the conditioning of the statistic obtained through the Fisher study.

In what follows, $WST_m$ refers to the concatenation of $S^{iso}_1$ and $\bar{S}^{iso}_2$ statistics for the $\psi_m$ wavelets, while $WST_w$ refers to the concatenation of $S_1$ and $\bar{S}_2$ for the $\psi_w$ wavelets. For each set of statistics, we thus have $J$ $S_1$ coefficients and $J\cdot (J-1)/2$ $S_2$ coefficients. This leads to 21 coefficients for $WST_m$ and 45 coefficients for $WST_m$, as $J_m =6$ and $J_w = 9$.

\subsection{Summary of evolution}
\label{Summary_of_Evolution}

In the previous section, except for the 3D power spectrum, all statistics are 2D and can be estimated on fixed-$z$ spatial slices. In this work, we want to include the evolution of our 2D coefficients along the lightcone. This means that these 2D statistics have to be estimated for each frequency channel, giving spatial 2D statistics with a $z$ dependency $\phi^s(z)$, which we call 2+1 statistics, where the $s$ labels the set of statistics (PS, WM, WST$_m$, WST$_w$).

\begin{table}
\centering
\caption[]{Change in number of coefficients before and after LoS summary.}
\begin{tabular}{l|||l|l}
Statistic&No. Coefficients&No. Coefficients\\
&Before LoS Summary &After LoS Summary\\
&&(Maximum)\\
    \hline\hline
$\phi_{PS}(z)$&1152&72\\
\hline
$\phi_{WM}(z)$&2304&144\\
\hline
$\phi_{WST_m}(z)$&2688&168\\
\hline
$\phi_{WST_w}(z)$&5760&360\\
\hline 
\end{tabular}
\begin{minipage}{\linewidth}
\textbf{Note:} As an example, for the LoS summary, we have assumed the maximum number of scales, \textbf{$j_z$} = 7.
\end{minipage}
\label{tab:Stats_Num}
\end{table}
As we work with lightcones with 128 frequency channels (see Sect.~\ref{Simulation_Set_Up}), the total number of coefficients for our 2+1 statistics is very high, as shown in Table ~\ref{tab:Stats_Num}. We thus chose to compress this evolution along the line of sight (LoS). To do this, we considered the application of a continuous wavelet transform\footnote{\url{https://pyWavelets.readthedocs.io/en/latest/ref/cwt.html}} to the evolution of the coefficient along the lightcone via the following cosine\footnote{In pyWavelets this is referred to as (real) Morlet wavelets.} wavelet:
\begin{equation}\label{eq:Morlet_Wavelet}
\psi_{\mathbf{j_z}}(t) = e^{-\frac{t^2}{2^{2\mathbf{j_z}}}}\cos\left(\frac{5t}{2^\mathbf{j_z}}\right).
\end{equation}
As these wavelets are constructed from a dyadic scaling, the maximum number of $j_z$ scales on which we can decompose these evolutions is $\log_{2}(128)=7$. 

Once we apply the cosine wavelets to the redshift evolution of each coefficient, we summarise their evolution by computing integrated $\ell^1$-and square $\ell^2$-norm:
\begin{equation}\label{eq:full_notation}
\bar{\phi}^s_{l_1:j_z} = || \phi^{s}(z) * \psi_{j_z}(z)||_1, ~~~~\text{and} ~~~~
\bar{\phi}^s_{l_2:j_z} = || \phi^{s}(z) * \psi_{j_z}(z)||_2^2.
\end{equation}
If we also include the mean of a coefficient's evolution, in addition to the maximum number of scale $j_z$ = 7 scales, we have reduced the number of coefficients by a factor 16; see Table~\ref{tab:Stats_Num}. This reduction in coefficients makes our resultant statistic far more stable. We note that this choice of wavelet compression arises from testing various methods, such as cosine decomposition and principal component analysis (PCA), to name just two, and we find the cosine wavelet to be the most stable and interpretable.

The $\ell^1$-norm and $\ell^2$-norm capture different evolution aspects of a given coefficient. Typically, the $\ell^2$-norm captures the energy or magnitude of the coefficients, while the $\ell^1$-norm captures sparsity and the presence of localised structures \citep{2011:BachJenattonMairal}. In the case where we summarise the evolution with the $\ell^1$-norm {and} square $\ell^2$-norm, one incorporates both the energy and sparsity characteristics of the coefficients, which leads to a more comprehensive representation of the underlying signal. This combination should provide an informative summary of the evolution across the lightcone. We refer to our `evolution-compressed' statistics as the evolution-compressed power spectrum (ecPS); the evolution-compressed wavelet moments (ecWM); the evolution-compressed WST$_m$ (ecWST$_m$), and the evolution-compressed WST$_w$ (ecWST$_w$). 

We note that in the following, we concatenate only part of these evolution compressed coefficients, for which we use short-hand notation. For instance, $\bar{\phi}_{\ell^1,\ell^2:1,2}^{WM}$ stands for the ec-WM computed at $j_z=1$ and $j_z=2$ scales for both $\ell^1$ and $\ell^2$ norms.

\section{Simulation setup} 
\label{Simulation_Set_Up}

\subsection{{\fontfamily{cmtt}\selectfont
21CMFAST}}

The {\fontfamily{cmtt}\selectfont
21CMFAST} \citep{2011:MesingerFurlanettoCen} code is a widely available semi-analytical tool for simulating the 21cm signal during the EoR\footnote{https://21cmfast.readthedocs.io/en/latest/}; it offers a computationally efficient alternative to hydrodynamic simulations by employing approximations for various physical processes. The simulation creates a high-resolution density field at a redshift of z=300. This field is then evolved to later redshifts using second-order Lagrangian perturbation theory \citep{1998:Scoccimarro}, which informs the spatial variations of the galaxy field through the conditional halo mass function. Dark matter halo fields are assigned galaxy properties through empirical scaling relations in accordance with the parametrisation described in \citet{2019:ParkMesingerGreig}.
{\fontfamily{cmtt}\selectfont
21CMFAST} employs an excursion-set-based formalism to identify ionised regions within the density field. This method involves considering spheres of decreasing radius, starting from a maximum radius $R_{\rm{max}}$, and determining if the number of ionising photons within the sphere exceeds the number of baryons. According to this formalism, the central pixel in a region is considered ionised if the following condition is satisfied:
\begin{equation}
\zeta f_{\rm{coll}} \geq 1,
\end{equation}
\newline
where $f_{\rm{coll}}$ is the fraction of collapsed matter within the spherical region under consideration. The maximum radius $R_{\rm{max}}$ represents the farthest distance a photon can travel within the simulated field before encountering a recombined atom. It is loosely related to the characteristic mean free path. The ionising efficiency of galaxies $\zeta$, is defined as:

\begin{equation}
\zeta=30\left(\frac{f_{\mathrm{esc}}}{0.3}\right)\left(\frac{f_{\star}}{0.05}\right)\left(\frac{N_\gamma}{4000}\right)\left(\frac{2}{1+n_{\mathrm{rec}}}\right)
,\end{equation}
where $f_{\text {esc}}$ represents the proportion of ionising photons that can escape into the intergalactic medium, $f_{\star}$ denotes the fraction of gas in a galaxy that is converted into stars, $N_\gamma$ stands for the quantity of ionising photons generated per baryon within stars, and $n_{\text {rec}}$ indicates the typical number of times a hydrogen atom undergoes recombination.

The 21cm brightness temperature map is calculated from the ionisation field using the following expression:
\begin{align}
\label{eq:Tbfinal}
\delta T_b &= 28(1+\delta)x_{\rm{HI}}\left(1-\frac{T_{\rm{CMB}}}{T_S}\right)\left(\frac{\Omega_bh^2}{0.0223}\right) \nonumber \\
&\quad \times \sqrt{\left(\frac{0.24}{\Omega_m}\right)\left(\frac{1+z}{10}\right)} \left[\frac{H}{\delta_rv_r + H}\right],
\end{align}
\newline
where $x_{\rm{HI}}$ denotes the neutral hydrogen fraction, $\delta$ represents the matter overdensity, $\delta_rv_r$ corresponds to the gradient of the line-of-sight peculiar velocity, $H(z)$ represents the Hubble parameter, and $T_{\rm{CMB}}$ and $T_S$ are the CMB and spin temperatures, respectively. The term $\delta_rv_r$ accounts for the impact of redshift space distortions, primarily affecting small-scale overdensities. For further details of the simulations, we refer the reader to \citet{2019:ParkMesingerGreig} and \citet{2020:MurrayGreigMesinger}.

\subsection{SKA noise}

In order to accurately simulate the thermal noise of the SKA, we use the \textsc{pstools}\footnote{\url{https://gitlab.com/flomertens/ps_eor/-/wikis/Radio-interferometer-sensitivty}} package. 
Our simulation methodology involves simulating the $uv$-coverage of SKA-Low. This means we simulate the $uv$ tracks for each individual baseline and then grid them into a $u v$ grid. The grid for $uv$-coverage is simulated based on the proposed distribution plan\footnote{\url{ https://www.skao.int/sites/ default/files/documents/d18- SKA- TEL- SKO- 0000422_02_SKA1_ LowConfigurationCoordinates-1.pdf}} for antennae in SKA-Low.

First, we must create a realistic 10 hour observation of the EoR-0 field ---a proposed SKA field located at an RA of $0.00\mathrm{h}$ and Dec $=-27 \mathrm{deg}$. To do this, we simulated a series of $uv$-tracks corresponding to this observation, with an integration time $t_{\text {int }}$ set at $10 \mathrm{s}$.
The noise within each $uv$-cell, and across each frequency channel $\nu$, was generated based on a Gaussian distribution. We calculated the standard deviation of this distribution using the formula:

\begin{equation}
  \sigma(u, v, \nu)=\frac{k_{\mathrm{B}} T_{\mathrm{sys}}(\nu)}{A_{\mathrm{eff}}} \sqrt{\frac{1}{2 \delta_v N(u, v, \nu) t_{\mathrm{int}}}}
.\end{equation}

Here, $k_{\mathrm{B}}$ is the Boltzmann constant, $T_{\mathrm{sys}}(\nu)$ is the system temperature at frequency $\nu$, $A_{\mathrm{eff}}$ represents the effective area, $N(u, v, \nu)$ denotes the number of baselines, $\delta_\nu$ signifies the frequency resolution, and $t_{\mathrm{int}}$ stands for the integration time.

This generates a $.h5$ which can be converted to images using the \textsc{pstools} function {\texttt{simu\_noise\_img}} ---this function can also be used to scale the number of observational hours from 10 to 100, as we have done in this work.

\subsection{Fisher analysis}
The Fisher matrix, introduced by Fisher in 1922 \citep{1922:Fisher}, is a powerful tool for estimating the accuracy with which a statistic can constrain a parameter.
Mathematically, the Fisher matrix is defined as follows:
\begin{equation} \mathbf{F_{ij}} = \bigg \langle \frac{\partial^2 \ln{\ell}}{\partial\theta_i\partial\theta_j}\bigg \rangle, \end{equation}
where $\ell$ represents the likelihood\footnote{Likelihood of the target signal given the model parameters and the sources of stochasticity (cosmic variance and thermal noise).} function; $\theta_i$ denotes the parameter being varied, and $\big<\big>$ represents the ensemble average, giving the expectation value. This matrix quantifies the sensitivity of the statistic to changes in the parameter. Under the assumption that the likelihood is a multivariate Gaussian, and thus the covariance matrix and mean vectors are sufficient information, we can further simplify the expression to \citep{Tegmark_1997,2013:Carron}:

\begin{equation}\label{eq:fisher} 
F^\theta_{ij} = \frac{\partial \textbf{S}}{\partial\theta_i} \mathbf{\Sigma}^{-1} \frac{\partial \textbf{S}}{\partial\theta_j}, 
\end{equation}
where \textbf{S} represents a vector encompassing the expected values of the statistics used, such as scattering coefficients or power spectrum $k$-bins, and $\mathbf{\Sigma}$ corresponds to the covariance of these statistics, which arises from cosmic variance or the thermal noise in high-noise cases. The covariance is estimated by conducting multiple independent simulation realisations at fixed parameter values, often chosen as fiducial values. To calculate the derivative, we introduced a small perturbation to the parameter $\theta_i$ around its fiducial value and performed multiple realisations with this perturbed parameter value.
According to the Cramer-Rao theorem, the variance of an unbiased estimator for a given parameter $\theta_i$ satisfies the following inequality:
\begin{equation} \delta^2\theta_i \geq \big(F^{-1}\big)_{ii}. \end{equation}
This Cramer-Rao bound establishes a lower bound on such a variance, indicating the smallest uncertainty achievable for an unbiased estimate of the parameter $\theta_i$.

\subsubsection{Fisher setup}
The simulated lightcone has a transverse extent of 200 $h^{-1} \rm{Mpc}$ and consists of 256 pixels per side in each frequency channel. The lightcone spans a redshift range from $z = 8.82$ (144.60 MHz) to $z = 9.33$ (137.46 MHz), comprising 128 frequency channels.

The three parameters we considered in our Fisher analysis are: $T_{Vir}$, which is minimum virial temperature needed for halos to host star-forming galaxies, as well as $R_{Max}$ and $\zeta$, as defined in the previous section. For each parameter, we varied them about the fiducial value:
\begin{enumerate}
  \item $T_{Vir}$: $50000\pm 5000$
  \item $R_{Max}$: $15\pm 5 Mpc$
  \item $\zeta$: $30\pm 5$,
\end{enumerate}
and ran 400 simulations for each parameter value; we then used these for the derivatives. For the covariance, we used 600 realisation simulations with fiducial astrophysical parameters.

\subsubsection{Fisher validation: Condition number}
When using the inverse of the covariance matrix, denoted $\mathbf{\Sigma}^{-1}$ in equation \eqref{eq:fisher}, it is important to ensure that the covariance matrix is well conditioned for numerical stability during the inversion process. The condition number\footnote{The condition number is a numerical measure used to assess how sensitive the output of a mathematical function or operation is to small changes in the input data. It is computed by dividing the largest eigenvalue of the covariance matrix by the smallest eigenvalue.} of the covariance matrix should ideally be of the order of or below 10$^7$ to be considered well conditioned \citep{2023:ParkAllysVillaescusa-Navarro}. If the covariance matrix is ill-conditioned, the inversion process can become unstable, leading to unreliable results in the Fisher formalism.

In Section \ref{Summary_of_Evolution}, we discussed different applications of equation \eqref{eq:lp_norm}, which involves the use of either $p=2$ (Euclidean norm) or a combination of $p=1$ and $p=2$. Though statistically limited to $j_z=7$, because
of the number of frequency channels, we must ensure that we are well-conditioned in the application of our statistics. For the $\ell^2$-norm, we find that we can use up to $j_z=4$ before encountering issues with conditioning. However, for the combination of the $\ell^1$-norm and $\ell^2$-norm, we can only use up to $j_z=2$ before encountering ill-conditioning.
\newline

We show a full description, including the number of coefficients, of all our statistics in Table \ref{tab:Stat_Description}. To avoid confusion between the different summaries, our evolution-compressed (ec) statistics ---ecPS, ecWM, ecWST$_m$, and ecWST$_w$ --- are referred to here using their notation shown in Table \ref{tab:Stat_Description}. 
\newline

\begin{table}

\centering
\caption[]{Description of Statistics}
\begin{tabular}{l|||l|l}
Label&Details&No.\\
&&Terms\\
    \hline\hline
$\phi^{PS}_{3D}$&Spherically-Averaged&9\\
&3D PS&\\
\hline
$\bar{\phi}_{\ell^2:1,2,3,4}^{PS}$&Statistic:2D PS (ecPS)&$9\times5$\\
&Summary: $\ell^2$-norm&\\
&Scales: $j_z$= 1,2,3,4&\\
\hline
$\bar{\phi}_{\ell^1,\ell^2:1,2}^{PS}$&Statistic:2D PS (ecPS)&$9\times5$\\
&Summary:$\ell^1$-norm + $\ell^2$-norm&\\
&Scales: $j_z$= 1,2&\\
\hline
$\bar{\phi}_{\ell^2:1,2,3,4}^{WM}$&Statistic:2D WM (ecWM)&$(9+9)\times5$\\
&Summary: $\ell^2$-norm&\\
&Scales: $j_z$= 1,2,3,4&\\
\hline
$\bar{\phi}_{\ell^1,\ell^2:1,2}^{WM}$&Statistic:2D WM (ecWM)&$(9+9)\times5$\\
&Summary:$\ell^1$-norm + $\ell^2$-norm&\\
&Scales: $j_z$= 1,2&\\
\hline
$\bar{\phi}_{\ell^2:1,2,3,4}^{WST_m}$&Statistic:2D WST$_m$ (ecWST$_m$)&$21\times5$\\
&Summary: $\ell^2$-norm&\\
&Scales: $j_z$= 1,2,3,4&\\
\hline
$\bar{\phi}_{\ell^1,\ell^2:1,2}^{WST_m}$&Statistic:2D WST$_m$ (ecWST$_m$)&$21\times5$\\
&Summary:$\ell^1$-norm + $\ell^2$-norm&\\
&Scales: $j_z$= 1,2&\\
\hline 
$\bar{\phi}_{\ell^2:1,2,3,4}^{WST_w}$&Statistic:2D WST$_w$ (ecWST$_w$)&$45\times5$\\
&Summary: $\ell^2$-norm&\\
&Scales: $j_z$= 1,2,3,4&\\
\hline
$\bar{\phi}_{\ell^1,\ell^2:1,2}^{WST_w}$&Statistic:2D WST$_w$ (ecWST$_w$)&$45\times5$\\
&Summary:$\ell^1$-norm + $\ell^2$-norm&\\
&Scales: $j_z$= 1,2&\\
\hline 
\end{tabular}
\begin{minipage}{\linewidth}
\textbf{Note:} Here we describe each statistic used in this work, including the number of coefficients that the application of the statistic will produce.   
\end{minipage}
\label{tab:Stat_Description}
\end{table}
Combining the first and second moments is crucial in order to accurately determine astrophysical parameters \citep{2019:AllysLevrierZhang,2022:GreigTingKaurov}. However, in our analysis of the statistics, combining $M_{1}(i)$ and $M_{2}(i)$ causes the covariance matrix to be poorly conditioned due to their values being of different orders of magnitude. In order to overcome this issue, we whitened the statistics by normalising them to their standard deviation. The condition numbers of the whitened statistics are presented in Table \ref{tab:Cond}. This normalisation balances the eigenvalues in the covariance matrix, making it better conditioned and leading to improved numerical stability. We applied this whitening technique to all our statistics, which did not affect the results of the Fisher analysis. This also proves that our covariances are stable, as applying whitening to an ill-conditioned matrix would change the results of the Fisher analysis.

In cases where there is no noise, we do not encounter any problems. We can observe that the noiseless statistics are well conditioned by referring to Table \ref{tab:Cond}, which displays the condition numbers of our various whitened statistics. However, when we introduce SKA noise into the data, we notice that the condition number for the noisy cases increases. This effect is not seen with the spherically averaged power spectrum; only with the 2+1 statistics. We find that understanding the results for 2+1 statistics is more difficult. Whether spatial statistics at various redshifts would exhibit more robust correlations concerning the EoR signal or the SKA noise remains to be seen. It is difficult to anticipate how the compression of $z$-information would impact the conditioning of the covariance matrix.

To ensure the numerical stability of the covariance, we performed an additional check by multiplying the inverse of the covariance by the covariance in order to verify that it is the identity matrix. Our findings reveal that the error on the off-diagonal terms ---which indicates how far they stray from zero--- is approximately $\sim10^{-8}$. Given that the condition numbers are of the order of $\sim10^{8}$, we would expect the error to be around $10^{-8}$ ---this is because double-precision computation allows for 16 digit representation--- signifying no significant instabilities. Despite the high condition number, the covariance matrices for high-noise cases are numerically stable. We present the errors on the off-diagonal terms for all the summary statistics in Table \ref{tab:err}.

The Fisher formalism is a valuable tool for estimating the information on parameters that is contained in a statistic. However, its effectiveness can be influenced by the difficulty in estimating ill-conditioned covariance matrices and derivatives. An alternative treatment would involve sampling the posterior distribution using a likelihood surrogate, which offers enhanced numerical stability and reliability in parameter estimations, making it a useful option. We use the Fisher matrix approach here for simplicity, but in future works, we will use more sophisticated approaches.

\begin{table}
\caption[]{Condition number of the statistics.}
\centering
\begin{tabular}{l|||l|l|l}
Statistics&No&100&1000\\
(Results are $\log_{10}$)&Noise&hours&hours\\[1ex]
    \hline\hline
$\phi^{PS}_{3D}$ &${4.46}$&${4.33}$&${4.28}$\\[1ex]
\hline
$\bar{\phi}_{\ell^2:1,2,3,4}^{PS}$ &${5.03}$&${6.52}$&${6.21}$\\[1ex]
\hline
$\bar{\phi}_{\ell^1,\ell^2:1,2}^{PS}$&${5.08}$&${6.49}$&${6.17}$\\[1ex]
\hline
$\bar{\phi}_{\ell^2:1,2,3,4}^{WM}$&${6.17}$&${8.10}$&${7.62}$\\[1ex]
\hline
$\bar{\phi}_{\ell^1,\ell^2:1,2}^{WM}$ &${5.47}$&${8.04}$&${7.6}$\\[1ex]
\hline
$\bar{\phi}_{\ell^2:1,2,3,4}^{WST_m}$ &${6.14}$&${9.03}$&${8.9}$\\[1ex]
\hline
$\bar{\phi}_{\ell^1,\ell^2:1,2}^{WST_m}$&${6.12}$&${8.98}$&${8.86}$\\[1ex]
\hline 
$\bar{\phi}_{\ell^2:1,2,3,4}^{WST_w}$ &${6.19}$&${7.66}$&${6.92}$\\[1ex]
\hline
$\bar{\phi}_{\ell^1,\ell^2:1,2}^{WST_w}$&${6.21}$&${7.56}$&${6.92}$\\[1ex]
\hline 
\end{tabular}
\begin{minipage}{\linewidth}
\textbf{Note:} To ensure the robustness of our statistical analysis, it is important to assess the condition numbers of the various statistics ---which have all been whitened--- employed in this paper under different noise treatments. A condition number below or of the order of $10^7$ is generally considered indicative of a well-conditioned covariance matrix.
\end{minipage}
\label{tab:Cond}
\end{table}

\begin{table}
\caption[]{Numerical stability of each statistic.}
\centering
\begin{tabular}{l|||l|l|l}
Statistics&No&100&1000\\
(Results are $\log_{10}$)&Noise&hours&hours\\[1ex]
    \hline\hline
$\phi^{PS}_{3D}$ &${-12.25}$&${-12.20}$&${-12.51}$\\[1ex]
\hline
$\bar{\phi}_{\ell^2:1,2,3,4}^{PS}$&${-11.87}$&${-9.24}$&${-10.14}$\\[1ex]

\hline
$\bar{\phi}_{\ell^1,\ell^2:1,2}^{PS}$&${-11.57}$&${-9.13}$&${-9.11}$\\[1ex]
\hline
$\bar{\phi}_{\ell^2:1,2,3,4}^{WM}$ &${-10.66}$&${-8.92}$&${-7.97}$\\[1ex]

\hline
$\bar{\phi}_{\ell^1,\ell^2:1,2}^{WM}$&${-11.20}$&${-8.64}$&${-9.14}$\\[1ex]
\hline
$\bar{\phi}_{\ell^2:1,2,3,4}^{WST_m}$&${-10.28}$&${-7.17}$&${-10.03}$\\[1ex]
\hline
$\bar{\phi}_{\ell^1,\ell^2:1,2}^{WST_m}$ &${-10.08}$&${-7.73}$&${-7.38}$\\[1ex]
\hline 
$\bar{\phi}_{\ell^2:1,2,3,4}^{WST_w}$&${-10.23}$&${-9.16}$&${-9.70}$\\[1ex]
\hline
$\bar{\phi}_{\ell^1,\ell^2:1,2}^{WST_w}$ &${-10.40}$&${-8.88}$&${-9.80}$\\[1ex]
\hline 
\end{tabular}
\begin{minipage}{\linewidth}
\textbf{Note:} Looking at the numerical stability of the different statistics, we multiply the inverse of the covariance by the covariance, which should be an identity matrix. We show the largest off-diagonal term, as a test for numerical stability.
\end{minipage}
\label{tab:err}
\end{table}
\subsubsection{Fisher validation: Convergence}
Another important check in the Fisher analysis is to ensure that our Fisher matrix is fully converged. We can check our convergence by comparing the diagonals of the Fisher matrices with different numbers of the simulation realisations (or samples) used to calculate them: 
\begin{equation}\label{eq:Convergence}
  \texttt{Convergence} = \frac{F_{ii}(N_{Samples})}{F_{ii}(N_{Max})}.
\end{equation}
We have a large number of samples with which to check this convergence. We performed two checks. In the first, we kept the number of samples used to calculate the derivatives constant at 400 and varied the number of fiducial samples to calculate the covariance. In the second, we kept the number of fiducial samples to calculate the covariance constant at 600 samples and only varied the number of samples used for the different astrophysical parameter derivatives. 

We show all of our convergence plots in Appendix \ref{Convergence_Plots}. When looking for convergence, we look to have less than a 10$\%$ error in our convergence check ---which we indicate here via the red-shaded region. We look to have our convergences stably inside this region, that is for at least approximately 100 samples\footnote{If we used 399 samples out of 400 it would be within this 10$\%$ region. Once within the 10$\%$ region, we can only be sure of convergence if we remain within this region as we increase the number of samples. Hence by seeing if the Fisher matrix in this region for a greater number of samples used, we can say that it has converged.}. We find that all of our Fisher matrices ---that is, for all statistics and noise cases--- are converged. 

\section{Results}
\label{Results}
We then carried out a Fisher analysis on our different statistics. We show corner plots comparing the best applications of our different statistics to the different noise cases of our data data, including $\phi^{PS}_{3D}$ as is traditionally used in 21cm parameter inference. We show all of the statistics in the Cramer-Rao bounds of this work for each parameter in this study in tables.
\citet{2022:GreigTingKaurov} previously compared wavelet scattering transforms to the power spectrum. In their work, the authors simulated a {\fontfamily{cmtt}\selectfont 21CMFAST} lightcone within the redshift range of 5.9 $\leq z \leq$ 27.4. They broke the lightcone into 12 chunks, applying the spherically averaged power spectrum to each chunk and applied the 2D WST to the central frequency slice of each light-cone chunk. 

In this work, we looked to simulate our lightcones in a single redshift band and at both the SKA angular resolution and frequency resolution. In addition to the $\phi^{PS}_{3D}$ statistic, we also used equation \eqref{eq:Morlet_Wavelet} to summarise the line-of-sight evolution of our different 2D statistics. This line-of-sight information can further constrain parameters.

\subsection{Noiseless}
\label{Noiseless_case}
 \begin{figure*}[!]
    \centering
    \begin{subfigure}[b]{\textwidth}
      \centering
      \includegraphics[width=.75\textwidth]{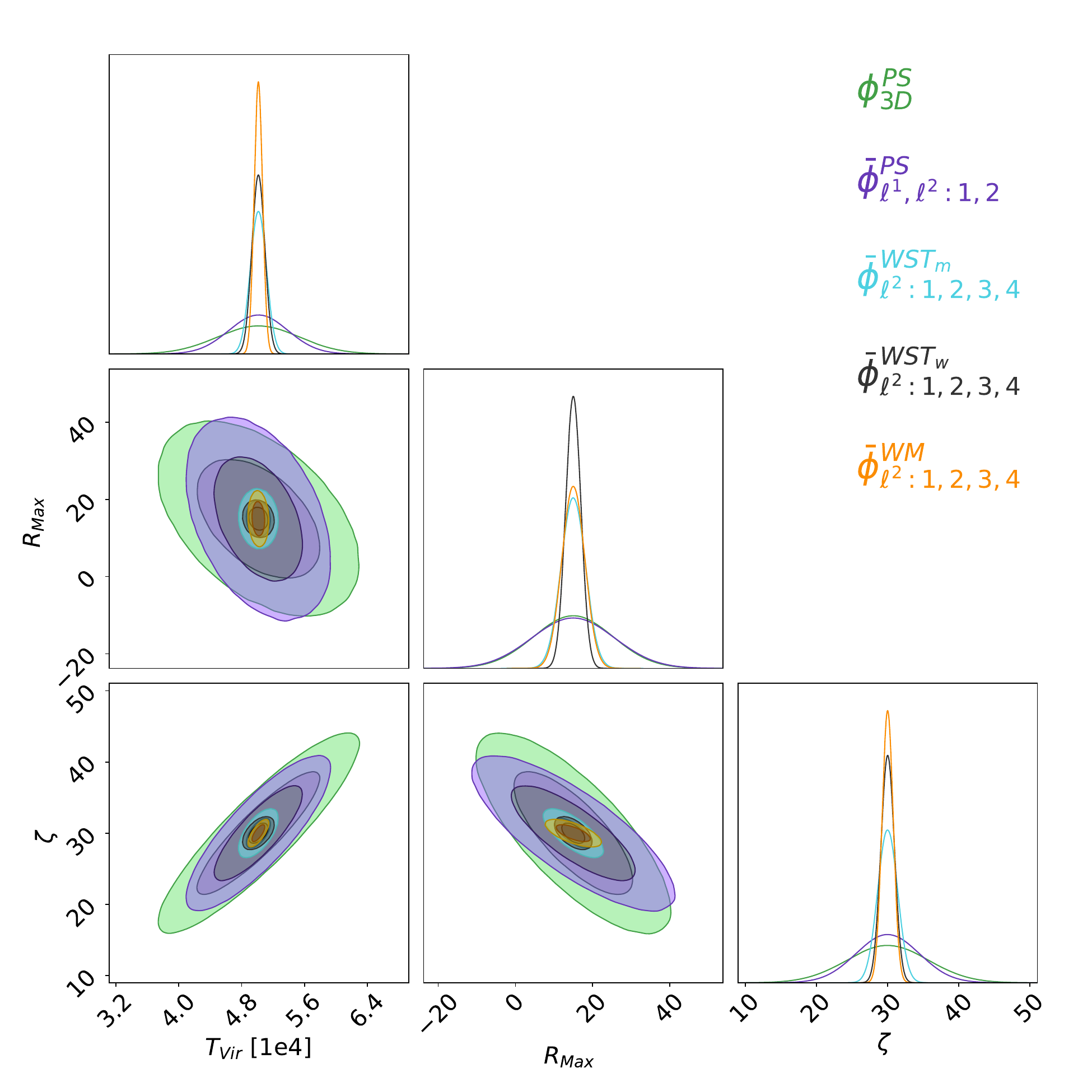}
    \end{subfigure}
    \begin{subfigure}[b]{\textwidth} 
      \centering 
      \includegraphics[width=\textwidth]{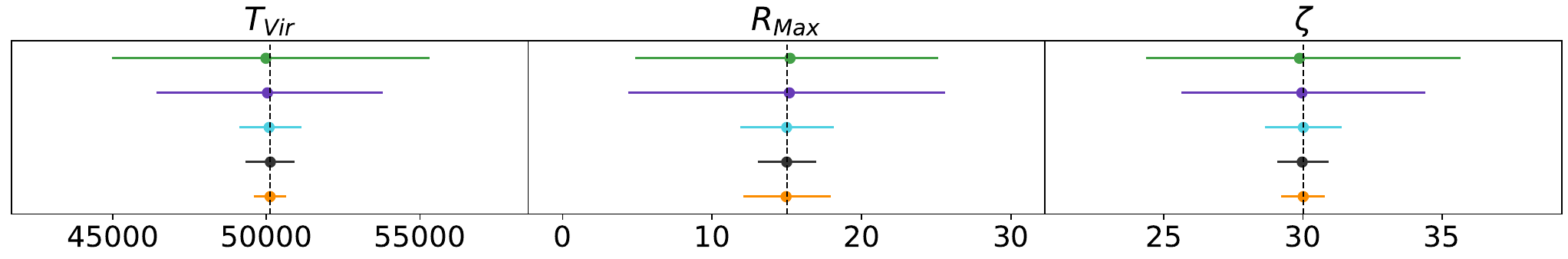}
    \end{subfigure}
    \caption{Results from the Fisher analysis of the three different summary statistics of the 21cm signal, when three astrophysical parameter are varied, as we have noiseless data, considering cosmic variance as the only source of variance. 
    \textit{Top:} Corner plot of our noiseless Fisher analysis, showing that $\bar{\phi}_{\ell^2:1,2,3,4}^{WM}$ provides the tightest contours.
    \textit{Bottom:} The $\pm 68\%$ credibility intervals of our different astrophysical parameters, for each statistic. 
    The ordering of the statistics is based on their performance, going from least constraining statistic (top) to most constraining (bottom).}
    \label{fig:WM_PS_No_Noise_Comparison_WF}
  \end{figure*}

From Fig.~\ref{fig:WM_PS_No_Noise_Comparison_WF} and Table \ref{tab:Bounds_no_noise}, we see that overall, $\bar{\phi}_{\ell^2:1,2,3,4}^{WM}$ produces the tightest constraints on the astrophysical parameters in our contour plots, closely followed by $\bar{\phi}_{\ell^2:1,2,3,4}^{WST_{w}}$ and $\bar{\phi}_{\ell^1,\ell^2:1,2}^{WST_{w}}$.
In their analysis of 2D WST versus the 3D power spectrum, \citet{2022:GreigTingKaurov} found that the WST produced slightly more precise constraints. Comparing Table \ref{tab:Bounds_no_noise}, we see that all the evolution-compressed statistics outperform the 3D power spectrum. Additionally, our wavelet-based statistics produce the tightest Cramer-Rao bounds. This is expected, as they combine the richer 2D information, such as that in the scattering transform as seen in \citet{2022:GreigTingKaurov}, while explicitly decomposing the line-of-sight information.

The evolution of the power spectrum along the lightcone exhibits high sensitivity to parameter changes, with the large-scale modes of the power spectrum being the most sensitive. Performing spherical averaging results in a loss of sensitivity to this line-of-sight information, as we mix both perpendicular and line-of-sight information. It is worth noting, however, that these conclusions are made for a single redshift band. 

For each of our 2+1 statistics, we performed two LoS decompositions: one summarises the evolution with the $\ell^2$-norm on scales $j_z = 1,2,3,4$ and the other with the $\ell^1$-norm + $\ell^2$-norm on scales $j_z = 1,2$. From Table \ref{tab:Bounds_no_noise}, it is difficult to make direct comparisons between the two LoS decompositions given that they operate on different scales. However, it should be noted that $\ell^2$-norm on scales $j_z = 1,2,3,4$ is expected to perform better than $\ell^2$-norm on scales $j_z = 1,2$. The similarity in the performance of the two decompositions shown in Table \ref{tab:Bounds_no_noise} suggests that $\ell^1$-norm + $\ell^2$-norm is more informative than $\ell^2$-norm, given that it takes two scales rather than four to produce comparable results. We are only restricting ourselves to small scales in the redshift domain ---the $\ell^1$-norm with $\ell^2$-norm only look at scales of size $2^1$ and $2^2$ --- rather than looking for large-scale information on the evolution. As mentioned above, we tried to include this large-scale information, but when we do, the condition number of our covariance matrix becomes too high. 

The contours in Fig.~\ref{fig:WM_PS_No_Noise_Comparison_WF} show that the astrophysical parameter constraints derived from the wavelet-based statistics and power spectra present similar degeneracies. Specifically, the $S_1^{iso}$ information alone from the WST$_m$, and the first layer of WST$_w$, are akin to that obtained from the power spectrum \citep{2019:AllysLevrierZhang,2020:ChengTingMenard}. However, applying WST$_m$ (and WST$_w$) here introduces an additional coefficient\footnote{Though we restricted ourselves to the isotropic RWST components, there also exist anisotropic components.} from $S^{iso1}_2$, which helps further constrain the parameters. Notably, although we still observe comparable degeneracies with the power spectrum, the degeneracy between $R_{Max}$ and $T{vir}$ is somewhat alleviated.

\begin{table}[!]
\caption[]{No Noise case, Cramer-Rao bounds.}
\centering
\begin{tabular}{l|||l|l|l}
Statistics&$T_{Vir}$&$R_{Max}$&$\zeta$\\
(Results are $\log_{10}$)&&\\
    \hline\hline
$\phi^{PS}_{3D}$ &${7.42}$&${2.01}$&${1.50}$\\[1ex]
\hline
$\bar{\phi}_{\ell^2:1,2,3,4}^{PS}$&${7.25}$&${2.05}$&${1.37}$\\[1ex]
\hline
$\bar{\phi}_{\ell^1,\ell^2:1,2}^{PS}$&${7.13}$&${2.04}$&${1.28}$\\[1ex]
\hline
$\bar{\phi}_{\ell^2:1,2,3,4}^{WM}$ &${5.45}$&${0.93}$&${-0.22}$\\[1ex]
\hline
$\bar{\phi}_{\ell^1,\ell^2:1,2}^{WM}$&${6.25}$&${1.00}$&${0.43}$\\[1ex]
\hline
$\bar{\phi}_{\ell^2:1,2,3,4}^{WST_m}$&${6.00}$&${0.98}$&${0.27}$\\[1ex]
\hline
$\bar{\phi}_{\ell^1,\ell^2:1,2}^{WST_m}$&${5.99}$&${1.01}$&${0.30}$\\[1ex]
\hline
$\bar{\phi}_{\ell^2:1,2,3,4}^{WST_w}$&${5.81}$&${0.58}$&${-0.07}$\\[1ex]
\hline
$\bar{\phi}_{\ell^1,\ell^2:1,2}^{WST_w}$&${5.79}$&${0.60}$&${-0.05}$\\[1ex]
\hline 
\end{tabular}
\begin{minipage}{\linewidth}
\textbf{Note:} Cramer-Rao bounds for all of our summary statistics in the case where we have no noise. The bound establishes a lower bound on the variance, i.e. the smallest uncertainty achievable for an unbiased estimate on a given parameter.
\end{minipage}
\label{tab:Bounds_no_noise}
\end{table}
\begin{table}[!]

\centering
\caption[]{100 hours of SKA Noise case, Cramer-Rao bounds.}

\begin{tabular}{l|||l|l|l}
Statistics&$T_{Vir}$&$R_{Max}$&$\zeta$\\
(Results are $\log_{10}$)&&\\
    \hline\hline
$\phi^{PS}_{3D}$ &${9.22}$&${3.60}$&${2.97}$\\[1ex]
\hline
$\bar{\phi}_{\ell^2:1,2,3,4}^{PS}$&${8.90}$&${3.52}$&${2.80}$\\[1ex]
\hline
$\bar{\phi}_{\ell^1,\ell^2:1,2}^{PS}$&${8.68}$&${3.24}$&${2.56}$\\[1ex]
\hline
$\bar{\phi}_{\ell^2:1,2,3,4}^{WM}$ &${8.16}$&${2.84}$&${2.27}$\\[1ex]
\hline
$\bar{\phi}_{\ell^1,\ell^2:1,2}^{WM}$&${8.09}$&${2.76}$&${2.20}$\\[1ex]
\hline
$\bar{\phi}_{\ell^2:1,2,3,4}^{WST_m}$&${9.31}$&${3.88}$&${3.21}$\\[1ex]
\hline
$\bar{\phi}_{\ell^1,\ell^2:1,2}^{WST_m}$&${9.22}$&${3.78}$&${3.12}$\\[1ex]
\hline  
$\bar{\phi}_{\ell^2:1,2,3,4}^{WST_w}$&${7.31}$&${1.73}$&${1.29}$\\[1ex]
\hline
$\bar{\phi}_{\ell^1,\ell^2:1,2}^{WST_w}$&${7.28}$&${1.71}$&${1.27}$\\[1ex]
\hline  
\end{tabular}
\label{tab:Bounds_100hrs_noise}
\end{table}

\subsection{One hundred hours of SKA noise}
\begin{figure*}[!]
    \centering
    \begin{subfigure}[b]{\textwidth}
      \centering
      \includegraphics[width=.75\textwidth]{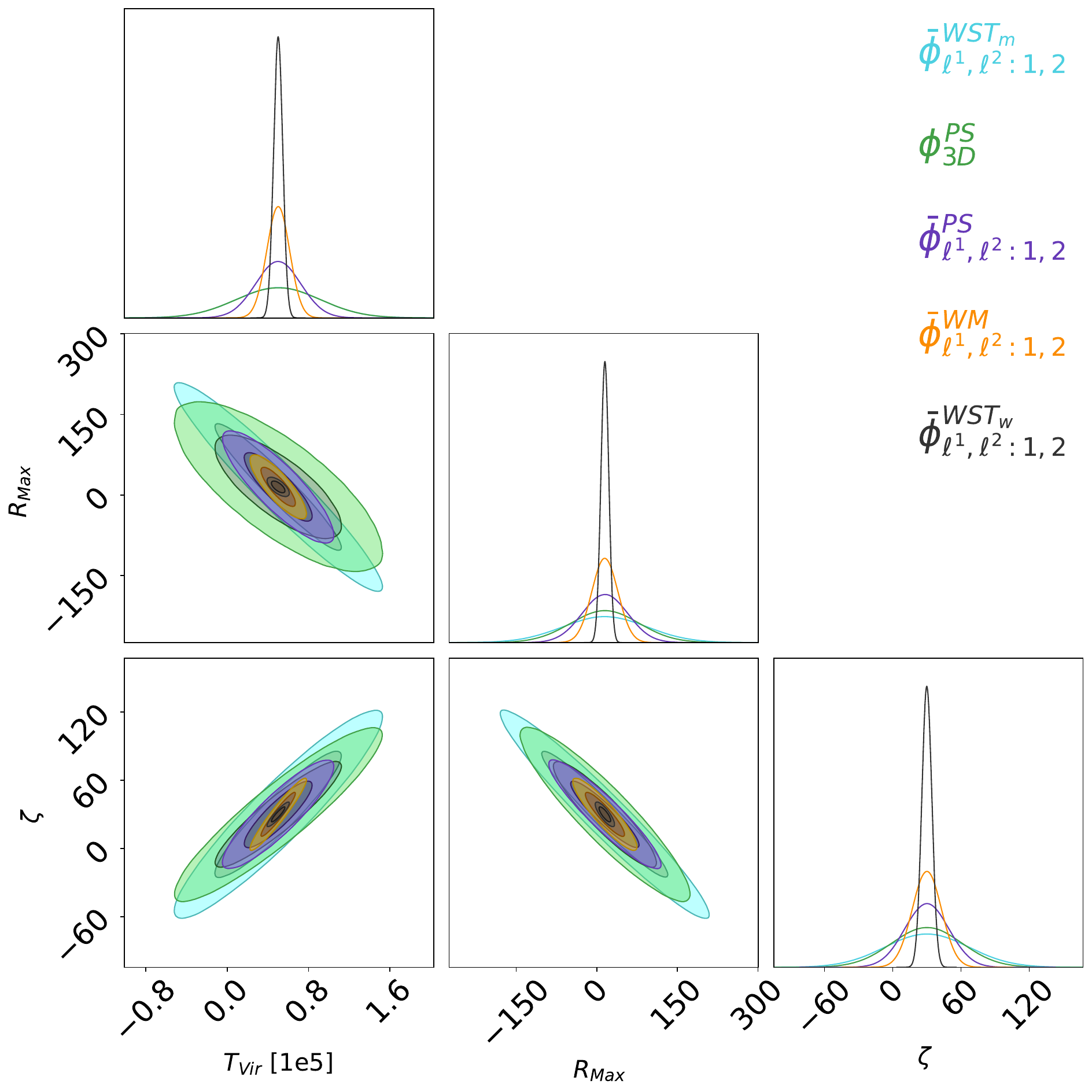}
    \end{subfigure}
    \begin{subfigure}[b]{\textwidth} 
      \centering 
      \includegraphics[width=\textwidth]{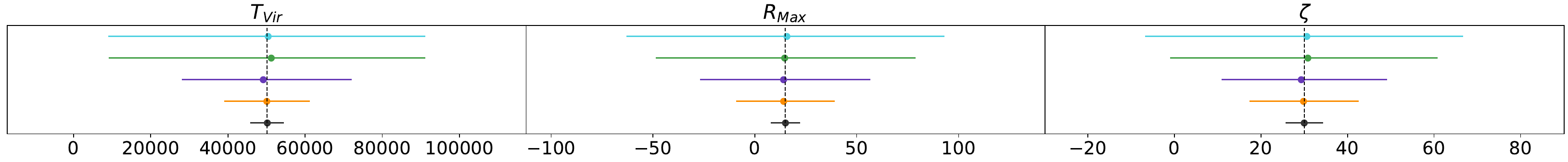}
    \end{subfigure}
    \caption{As in Fig.~\ref{fig:WM_PS_No_Noise_Comparison_WF}, but for 100 hours of SKA noise, where the noise is the dominant source of variance. We see now that WST$_m$ is the statistic with the worst performance, and that WST$_w$, with its evolution along the lightcone summarised by the $\ell^1$-norm and $\ell^2$-norm on scales $j_z=1,2$, which utilises wavelets derived from the power spectra binning, provides the tightest contours.}
    \label{fig:RWST_WM_PS_Comparison_100hr}
  \end{figure*}

We now look to see how robust these statistics are to thermal noise by comparing how well they constrain these parameters with the inclusion of two levels of noise. We start by looking at 100 hours of SKA observation level noise.
We show the Cramer-Rao bounds of each statistic and for each parameter in Table \ref{tab:Bounds_100hrs_noise}, and show the same subset of each statistic in Fig.~\ref{fig:RWST_WM_PS_Comparison_100hr}, as shown in Fig.~\ref{fig:WM_PS_No_Noise_Comparison_WF}. 

In the noiseless case, we see similar results between summarising the evolution of the statistics with either the $\ell^2$-norm alone or a combination of the $\ell^1$-norm and $\ell^2$-norm. Now we see in Table \ref{tab:Bounds_100hrs_noise} that for all of evolving statistics, the two scales summarised with the $\ell^1$-norm and $\ell^2$-norm outperform the four scales summarised with the $\ell^2$-norm. When there is noise in our data, the noise can cause extra variations and alter the magnitude characteristics of the coefficients. This means that relying solely on the $\ell^2$-norm may be more dependable, as there are a significant number of scales not affected by the noise. The $\ell^1$-norm, on the other hand, is able to extract more information from a smaller subset of scales, which could be less affected by the noise.

From Fig.~\ref{fig:RWST_WM_PS_Comparison_100hr}, we observe a notable contrast in the results compared to Figure \ref{fig:WM_PS_No_Noise_Comparison_WF}. Specifically, with $\bar{\phi}_{\ell^1,\ell^2:1,2}^{WST_w}$ producing the tightest constraints, $\bar{\phi}_{\ell^2:1,2,3,4}^{WST_m}$ is now the least constraining statistic, with not only $\bar{\phi}_{\ell^1,\ell^2:1,2}^{PS}$ producing tighter constraints but also $\bar{\phi}_{3D}^{PS}$. The $\bar{\phi}_{\ell^1,\ell^2:1,2}^{PS}$ statistic produces, on average, $15\%$ tighter parameter constraints compared to $\bar{\phi}_{\ell^2:1,2,3,4}^{WST_m}$, and $\bar{\phi}_{\ell^1,\ell^2:1,2}^{WM}$, on average, produces $40\%$ tighter parameter constraints. 

The power spectra are understood to outperform WST$_m$ ---with its current choice of wavelets--- in high-noise situations due to its averaging effect over multiple Fourier modes during binning as well as the fact that it probes the scale resolutions of interest, which enhances the signal-to-noise ratio. The binning strategy used by the power spectrum is more sensitive to large-scale structures, whereas thermal noise typically affects the smallest scales. Consequently, even with averaging, the power spectrum primarily captures structural information where noise is less likely to dominate.
In contrast, WST$_m$ coefficients here are predominantly associated with smaller scales, making them more susceptible to noise. The abundance of coefficients for smaller scales amplifies the influence of noise, which dominates at those scales. We find the three largest $k$-bins of the power spectra, which are also used by the WM application, to be severely noise dominated. These are associated with the scales $2^1$ and $2^2$ for WST$_m$, which contain a total of 55 evolved coefficients, roughly half of the number of coefficients for WST$_m$. 

For the WMs, we resolved this by leveraging the binning strategy of the power spectrum using the inverse Fourier transform of its binning window function as the wavelet of choice in order to exploit its spectral resolution to probe structural information where noise is unlikely to dominate.
As expected, the WMs outperform the 2D power spectrum. We see that including the $M_{1}(i)$ in the WMs introduces a statistical measure that exhibits greater sensitivity to parameter changes than noise.
We further demonstrate the importance of a more appropriate wavelet with the results of WST$_w$. For WST$_w$, we use a similar\footnote{As the wavelets used are isotropic, with no directional component.} scattering transform formalism but with the same wavelets as those used with the WMs. In this case, we benefit from the informative scattering transform on scales less affected by the noise, as now fewer of its components are in the less informative and noise-dominated bands, as they are with WST$_m$. 
For both LoS decompositions, WST$_w$ provides the tightest constraints (see Table \ref{tab:Bounds_100hrs_noise}) and the Cramer-Rao bounds are orders of magnitude better than those of the other statistics.

\subsection{One thousand hours of SKA noise}
\begin{table}[t]

\centering
\caption[]{1000 hours of SKA Noise case, Cramer-Rao bounds.}

\begin{tabular}{l|||l|l|l}
Statistics&$T_{Vir}$&$R_{Max}$&$\zeta$\\
(Results are $\log_{10}$)&&\\
    \hline\hline
$\phi^{PS}_{3D}$ &${8.82}$&${2.77}$&${2.24}$\\[1ex]
\hline
$\bar{\phi}_{\ell^2:1,2,3,4}^{PS}$ &${7.85}$&${2.74}$&${1.94}$\\[1ex]
\hline
$\bar{\phi}_{\ell^1,\ell^2:1,2}^{PS}$&${7.82}$&${2.53}$&${1.82}$\\[1ex]
\hline
$\bar{\phi}_{\ell^2:1,2,3,4}^{WM}$ &${7.56}$&${2.36}$&${1.70}$\\[1ex]
\hline
$\bar{\phi}_{\ell^1,\ell^2:1,2}^{WM}$&${7.50}$&${2.23}$&${1.62}$\\[1ex]
\hline
$\bar{\phi}_{\ell^2:1,2,3,4}^{WST_m}$&${8.10}$&${2.80}$&${2.00}$\\[1ex]
\hline
$\bar{\phi}_{\ell^1,\ell^2:1,2}^{WST_m}$&${8.11}$&${2.79}$&${2.00}$\\[1ex]
\hline 
$\bar{\phi}_{\ell^2:1,2,3,4}^{WST_w}$&${6.74}$&${1.24}$&${0.65}$\\[1ex]
\hline
$\bar{\phi}_{\ell^1,\ell^2:1,2}^{WST_w}$&${6.69}$&${1.24}$&${0.63}$\\[1ex]
\hline 
\end{tabular}
\label{tab:Bounds_1000hrs_noise}
\end{table}

\begin{figure*}[!]
    \centering
    \begin{subfigure}[b]{\textwidth}
      \centering
      \includegraphics[width=.75\textwidth]{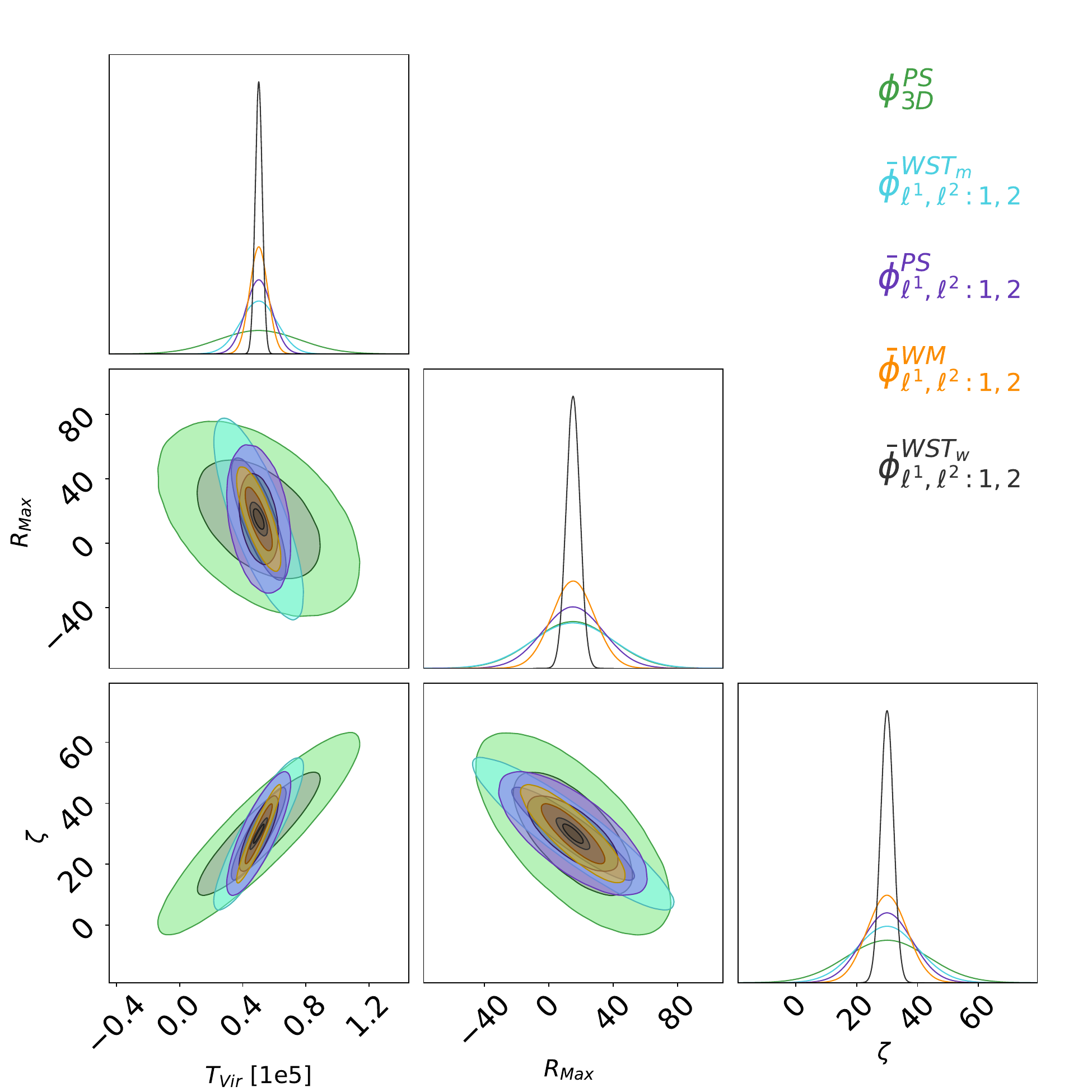}
    \end{subfigure}
    \begin{subfigure}[b]{\textwidth} 
      \centering 
      \includegraphics[width=\textwidth]{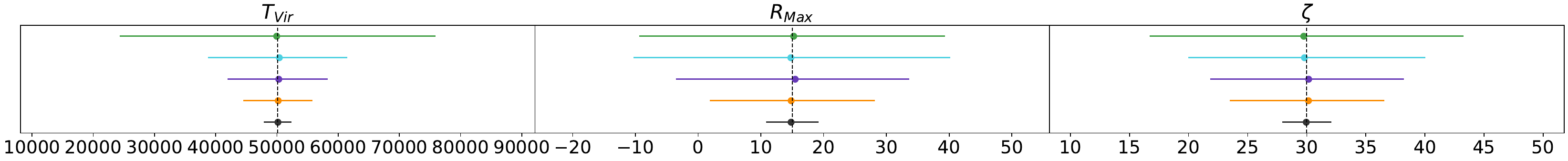}
    \end{subfigure}
    \caption{ Same as Fig.~\ref{fig:RWST_WM_PS_Comparison_100hr}, but now we have 1000 hours of SKA noise. WST$_m$ is still the worst performing evolution-compressed statistic in the low-noise case, but it is now closer to the precision of the power spectrum. The WST$_w$ statistic still provides the most precise constraints.}
    \label{fig:RWST_WM_PS_Comparison_1000hr}
  \end{figure*}

We now consider 1000 hours of SKA noise, and show Cramer-Rao bounds of the different statistics for each parameter in Table \ref{tab:Bounds_1000hrs_noise}. We show the best result of each statistic in Fig.~\ref{fig:RWST_WM_PS_Comparison_1000hr}, for 1000 hours of SKA noise.

In this low-noise regime, ecWST$_w$ still outperforms the other evolution-compressed statistics ---with both wavelet-based techniques using the wavelet derived from the power spectrum binning scheme performing the best. We see that the results of ecWST$_m$ have significantly improved in this lower-noise case. The precision difference between ecWST$_m$ and the other statistics has been significantly reduced in this case. In this lower-noise regime, the smaller scales are less impacted by the noise, allowing the coefficients of ecWST$_m$ to have a higher signal-to-noise ratio and a better statistical constraining power.

Whereas before the $\ell^1$ and $\ell^2$ combination produced the highest precision constraints for our evolution-compressed statistics, Table \ref{tab:Bounds_1000hrs_noise} shows that now, at least for the wavelet-based statistics, the $\ell^2$ LoS summary performs at least as well as the $\ell^1$ and $\ell^2$ summary. We also see that the 2+1 evolution-compressed statistics still produce constraints of higher
precision overall than $\phi^{PS}_{3D}$. However, for $R_{Max}$, we see that the posterior for $\phi^{PS}_{3D}$ produces tighter constraints than that for ecWST$_m$ for both LoS decompositions. We also see, for the first time, a significant change in the degeneracies of our statistics. For $R_{Max}$--$T_{Vir}$ and $\zeta$--$T_{Vir}$, we see that the two evolution-compressed statistics have a different degeneracy compared to that of $\phi^{PS}_{3D}$.
In real-world observations, there are no cases that are completely free of noise. We have indeed gained the advantages intended by choosing a set of wavelets more suited to the spectral domain of interest for our signal. In these noise cases, we are able to probe the important large scales whilst limiting the effect of noise and utilise the WMs to enhance our constraints.

\section{Conclusions}
\label{Conclusions}

In this paper, we present an approach to enhance the accuracy of astrophysical parameter constraints when analysing the 21cm signal from the epoch of reionisation (EoR). We focus on the use of 2+1 statistics, which include spectral evolution of the 2D spatial statistics.
In addition to comparing different summary statistics such as the power spectrum, WMs, and the WSTs, we also explore the importance of the choice of wavelet basis to improve sensitivity to astrophysical parameters. For WM and WST$_w$, we aim to use the binning strategy usually used for power spectra in EoR studies. This is expected to yield more accurate parameter estimations by choosing a frequency sampling relevant to the EoR signal while decreasing the contribution of noise to the summary statistics. By way of comparison, for WST$_m$ we retained the dyadic wavelet sampling used in previous astrophysical studies. We investigated the performance of these statistics in both noiseless and noisy scenarios, with noise levels corresponding to 100 and 1000 hours of observations with the SKA.

In the noiseless scenario, we discovered that all the 2+1 evolution-compressed statistics perform better than the spherically averaged power spectrum. The wavelet statistics outperform the ecPS, with ecWM ---with the $\ell^2$ LoS decomposition on redshift scales $j_z = 1,2,3,4$, and both ecWST$_w$--- providing the tightest limits of all the statistics. This shows that a simpler 2+1 statistical method can provide precise parameter constraints. These wavelet statistics can effectively identify non-Gaussian information beyond two-point statistics. WST$_m$ combines $S^{iso}_1$ and $S^{iso,1}_2$, while WST$_w$ combines the first and second layer ---as also shown by \citet{2022:GreigTingKaurov}. For WM, combining the $M_{2}(i)$ coefficients ---which are similar to the 2D power spectrum--- with the $M_{1}(i)$ coefficients allows us to investigate sparsity.

In both cases of noise, that is, the high-noise case (100 hours of SKA noise) {and the} low-noise case (1000 hours of SKA noise), overall, the 2+1 evolution-compressed statistics, including the ecPS, outperform the spherically averaged power spectrum. This suggests that the 2+1 wavelet-based statistics proposed in this paper form an effective alternative to the direct introduction of 3D statistics.

We also see that ecWST$_w$ far outperforms other statistics in these noisy regimes. This demonstrates that proper choices of wavelet and scale sampling are critical when building wavelet statistics. Indeed, the ecWST$_m$ under-performed here in high-noise cases due to its improper sampling of the largest scales, which are less contaminated by noise. Interestingly, both LoS summaries of ecWST$_w$ perform similarly, whereas the other statistics see a marked improvement in their constraints in high-noise cases by using the $\ell^1$ and $\ell^2$ combination for the LoS summary.
\newline
Real-world observations are subject to additional sources of nuisance and complexity. Various factors can affect the data, such as residual gain errors, imperfect excision of radio frequency interference (RFI), and other instrumental artefacts. Therefore, it is crucial to incorporate these elements into the analysis framework. This step is necessary to make astrophysical parameter estimation more robust and reflective of the challenges that arise in practical applications.

A word of caution
is warranted regarding the limitations of the Fisher approach. The Fisher approach, which approximates the likelihood by a single multivariate Gaussian at the fiducial point, simplifies the parameter space into ellipses; these poorly represent the true shapes of real parameter posteriors, which exhibit complex, often non-Gaussian shapes, reflecting the degeneracies and correlations between astrophysical parameters. Recognising this limitation should drive future research towards more sophisticated statistical methodologies that can capture the full complexity of the parameter space~\citep{2023:ParkAllysVillaescusa-Navarro}.

Finally, we restricted ourselves to the isotropic components of our data. In reality, the 21cm signal is highly non-Gaussian in nature, and the performance of the WST statistics would likely have been enhanced if we had included its anisotropic components (see \citet{2019:AllysLevrierZhang}). The results of this paper highlight the potential of 2+1 statistics and wavelet-based statistics for future 21cm intensity mapping studies in cosmology. We have created a GitHub repository\footnote{\url{https://github.com/ihothi/Wavelet_Stats}} containing the scripts used to produce the statistics in this paper, and detailing how to reproduce the simulations used in this work.

\begin{acknowledgements}

The post-doctoral contract of Ian Hothi was funded by Sorbonne Université in the framework of the Initiative Physique des Infinis (IDEX SUPER). We would like to thank Florent Mertens and Andrea Bracco for their useful discussions and input. We also thank our referee Prof. Yuan-Sen Ting for their comments to make the paper more balanced and their discussion points. This research made use of astropy, a community-developed core Python package for astronomy \citep{Astropy}; scipy, a Python-based ecosystem of open-source software for mathematics, science, and engineering \citep{Scipy} - including numpy \citep{numpy}; matplotlib, a Python library for publication quality graphics \citep{matplotlib}.

\end{acknowledgements}

\bibliography{cit} 
\appendix

\section{Fisher convergences}
\label{Convergence_Plots}

In this section, we analyse the convergence of different statistics using equation \eqref{eq:Convergence}. We use the Fisher matrix as a measure of convergence, which is calculated using a varying number of simulations. Our objective is to determine when convergence is achieved when the Fisher matrix calculated using a certain number of simulations is identical to the Fisher matrix calculated using the maximum number of simulations. To this end, we plot the convergence ratio between the Fisher matrix calculated for a given number of simulations and the one calculated using the maximum number of simulations. This plot helps us to determine the convergence region, which is the range where the ratio is considered to have converged within a 10\% error. We allow for a convergence error because complete convergence is challenging to achieve.
This task can be easily misinterpreted. As we increase the number of simulations to calculate the Fisher, we are approaching `convergence' of our Fisher matrix. For instance, if we vary the number of simulations used to calculate the derivatives, the ratio may appear within the convergence region at 250 simulations. One might assume that convergence has been achieved and stop there. However, at 255 simulations, the ratio may diverge from convergence. Therefore, when we perform such an analysis, we must ensure that the Fisher matrix is within the convergence zone for many simulations (used in calculations) after the supposed convergence. This approach helps us to confirm that we have truly reached convergence.

In Appendix \ref{Convergence_Plots_Cov}, we investigate the convergence of the number of fiducial simulations used to calculate the covariance for our fisher analysis. We keep the number of simulations used to calculate the derivatives fixed at 400. For each Fisher matrix calculated using a given number of simulations/samples, as shown in equation \eqref{eq:Convergence}, we divided it by the Fisher matrix using the total number of fiducial simulations. This ratio is plotted as a function of the number of simulations/samples.

In Appendix \ref{Convergence_Plots_Deriv}, we calculate the Fisher matrix, keeping the number of fiducial simulations fixed at 600 while varying the number of simulations or `samples' used to calculate the derivatives. We aim to plot the Fisher as a function of the number of simulations or samples. In each of our plots, we look for the error ---indicated with a red shaded region--- to be below $10\%$.

\subsection{Covariance convergence}
\label{Convergence_Plots_Cov}
\begin{figure*}[t]
\begin{subfigure}{.35\linewidth}
 \centering
 \includegraphics[width=\linewidth]{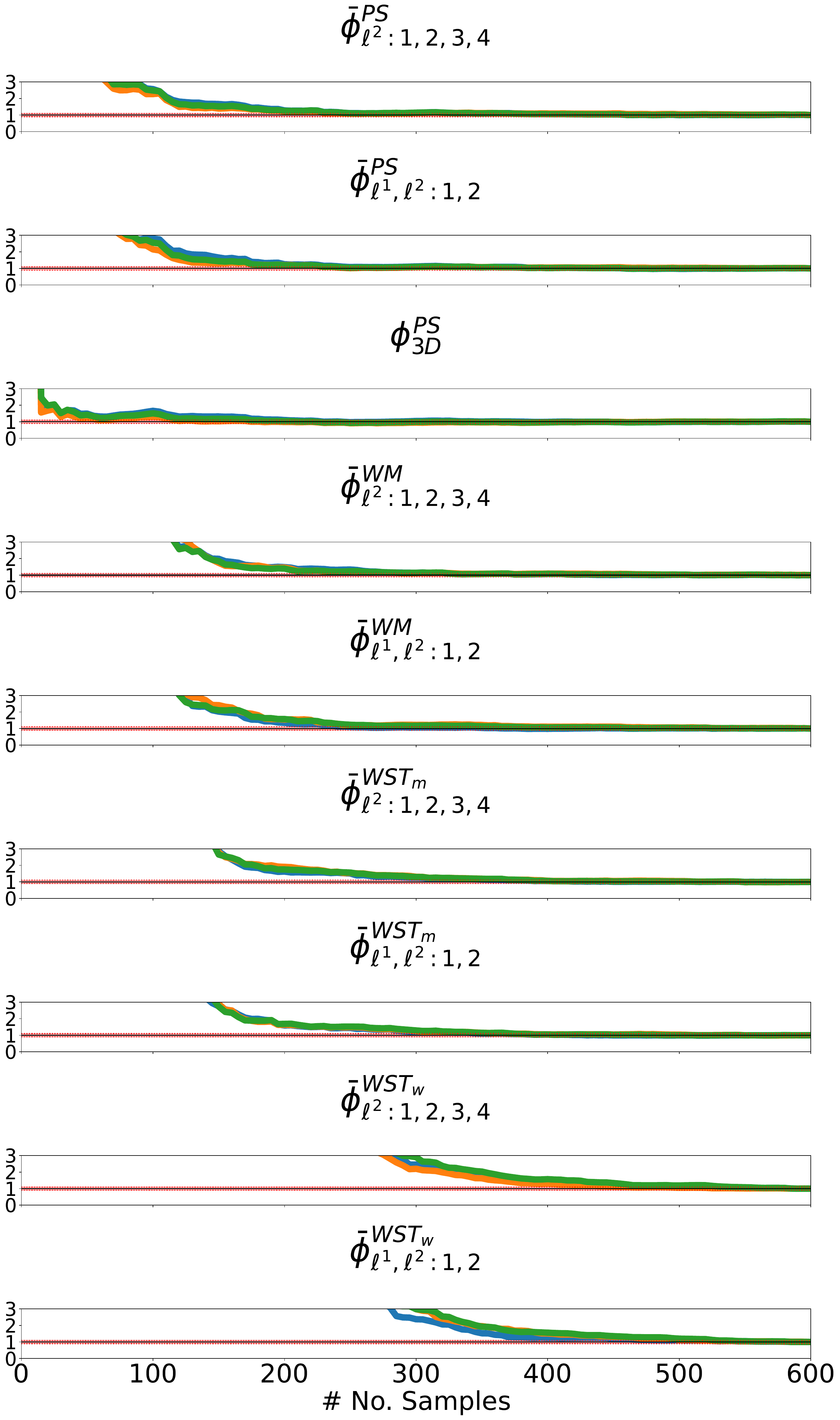}
 \caption{No Noise}
 \label{subfig:Full_Noiseless_convergence_cov_vary}
\end{subfigure}%
\begin{subfigure}{.35\linewidth}
 \centering
 \includegraphics[width=\linewidth]{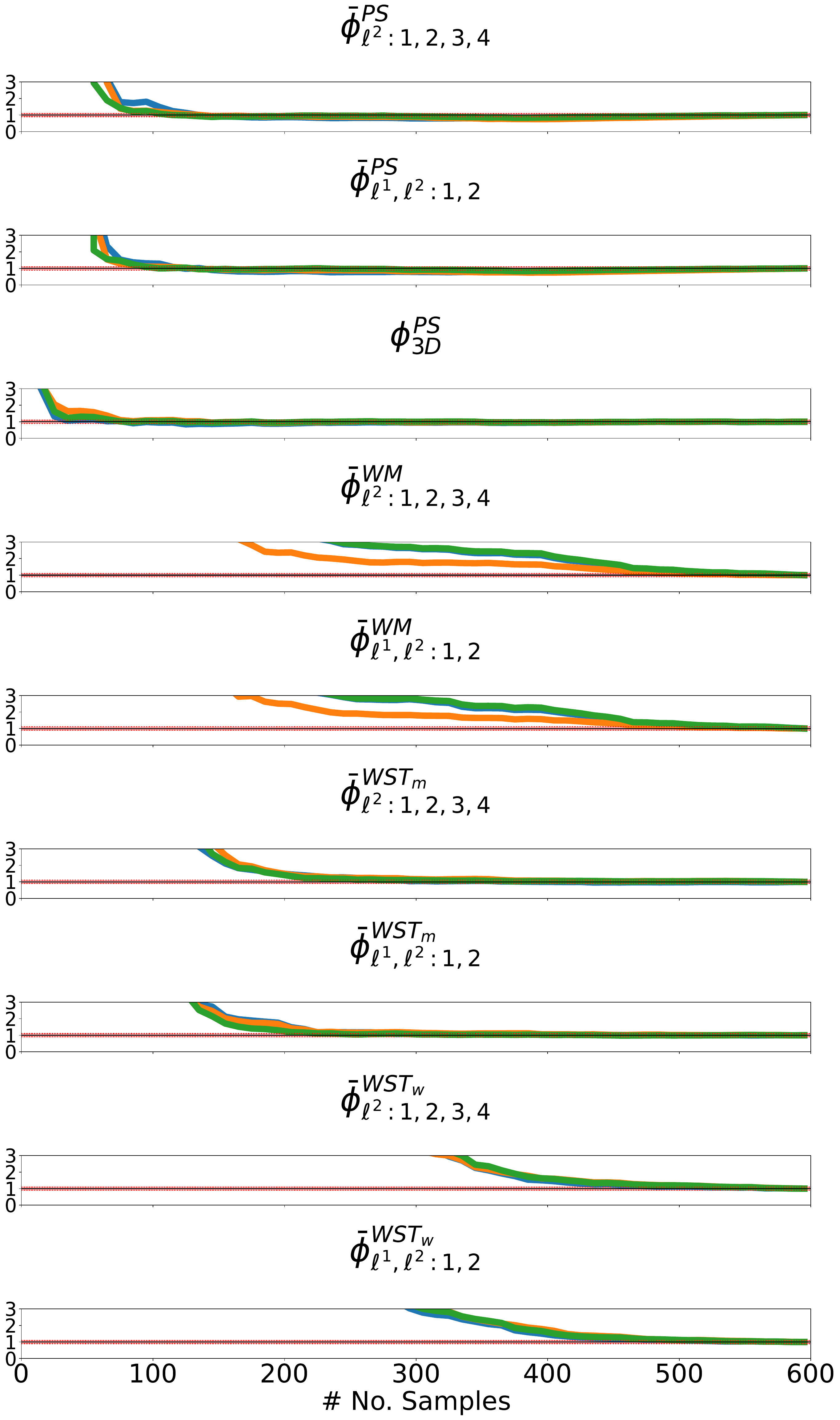}
 \caption{100 hours of SKA noise}
 \label{subfig:Full_Noiseless_convergence_100hrs_cov_vary}
\end{subfigure}%
\begin{subfigure}{.35\linewidth}
 \centering
 \includegraphics[width=\linewidth]{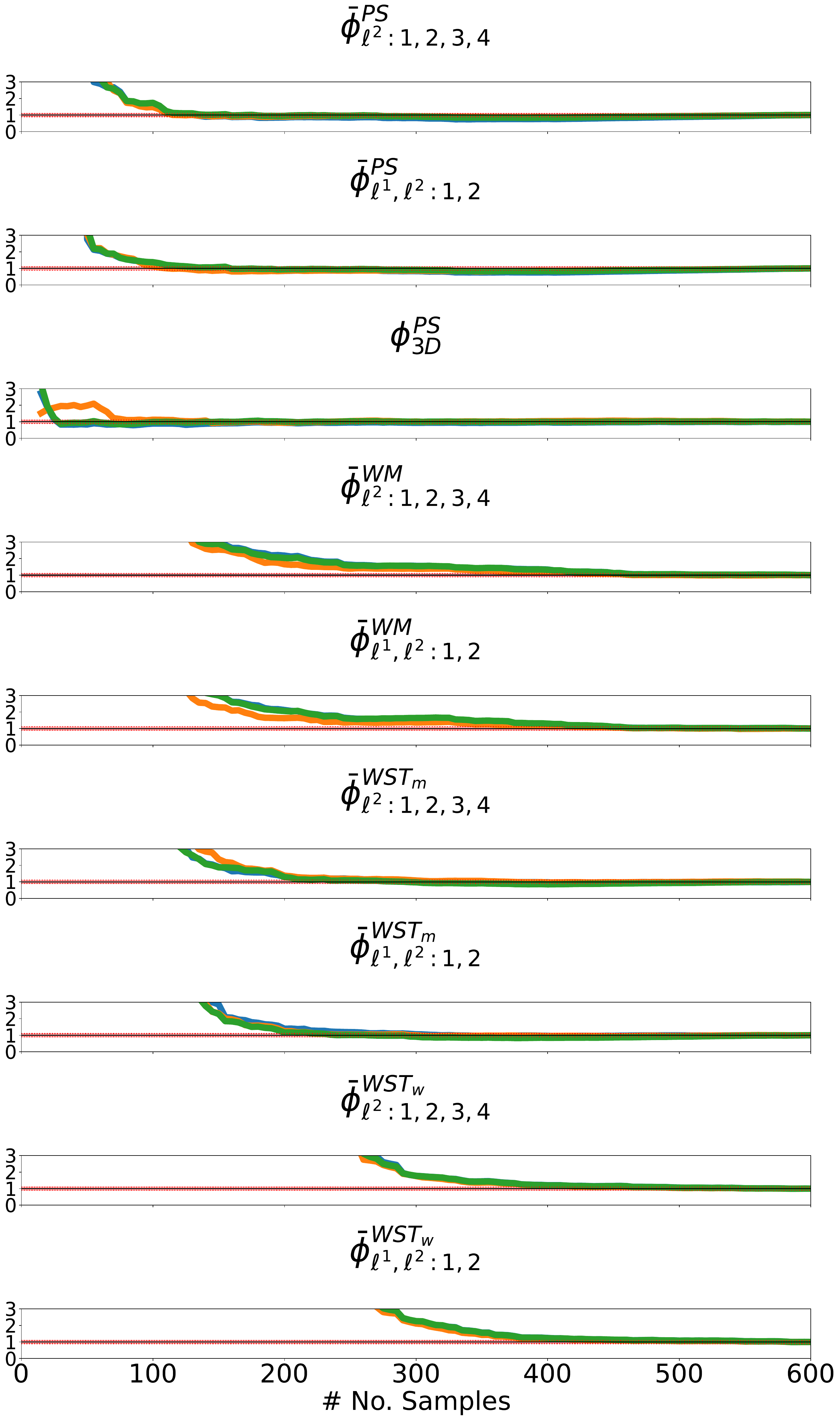}
 \caption{1000 hours of SKA noise}
 \label{subfig:Full_Noiseless_convergence_1000hrs_cov_vary}
\end{subfigure}
\caption{Convergence plots using equation \eqref{eq:Convergence}, where we have kept the number of simulations used for our derivatives constant at 400 and are solely varying the number of fiducial simulations used for the covariance. We consider the different noise cases: (a) no noise, (b) 100 hours of SKA noise, and (c) 1000 hours of SKA noise. We see that by 200 samples all of our statistics are fully convergent, falling within the 10$\%$, shown as the red shaded region. }\label{fig:Convergence_plots_cov_vary}
\end{figure*}

In this test, we keep the number of simulations (or samples) used to calculate the derivatives constant at 400, and vary the number of fiducial simulations used to calculate the covariance from 1 to 600.
We can see from Fig.~\ref{subfig:Full_Noiseless_convergence_cov_vary} that most of our different statistics are fully converged by 300 samples. We see that our evolution-compressed statistics require at least 300 samples before converging, whilst the spherically averaged power spectrum is convergent after 200 samples. WST$_w$ needs around 400 samples in the no-noise case before it becomes fully convergent.
In Fig.~\ref{subfig:Full_Noiseless_convergence_100hrs_cov_vary}, we show the convergence plots of our statistics in a high-noise scenario. Whereas the noiseless case will demonstrate the point of convergence with respect to cosmic variance, the noisy cases will probe convergence of our noise realisations. We see that most of the statistics reach convergence after 200 samples. The power spectrum statistics converge most readily, converging after 150 samples, whereas most of the wavelet-based statistics need a few hundred more samples to be convergent. WST$_w$ requires close to 450 samples before it is fully convergent. 
The convergence plots for our statistics under 1000 hours of SKA noise are presented in Fig.~\ref{subfig:Full_Noiseless_convergence_1000hrs_cov_vary}. We observe a similar trend to that in Fig.~\ref{subfig:Full_Noiseless_convergence_100hrs_cov_vary}, where noise has a significant impact and we need approximately 200 samples to achieve noise convergence. 
We can see that our wavelet-based statistics typically require more samples than the power-spectra-based statistics before they are fully convergent. This is likely due to the non-Gaussian information they contain, requiring more samples.

\subsection{Derivative convergence}
\label{Convergence_Plots_Deriv}
\begin{figure*}[t]
\begin{subfigure}{.35\linewidth}
 \centering
 \includegraphics[width=\linewidth]{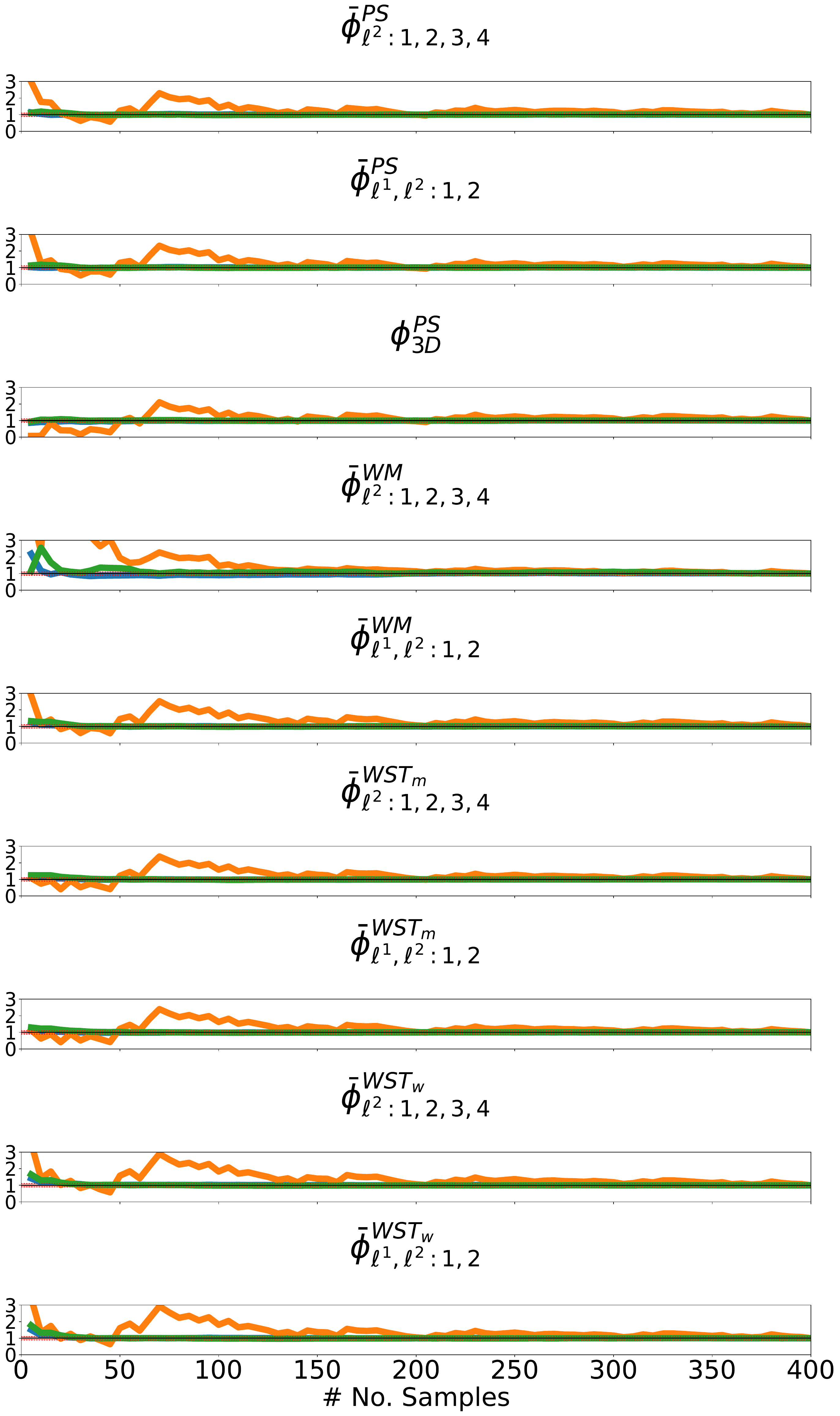}
 \caption{No Noise}
 \label{subfig:Full_Noiseless_convergence}
\end{subfigure}%
\begin{subfigure}{.35\linewidth}
 \centering
 \includegraphics[width=\linewidth]{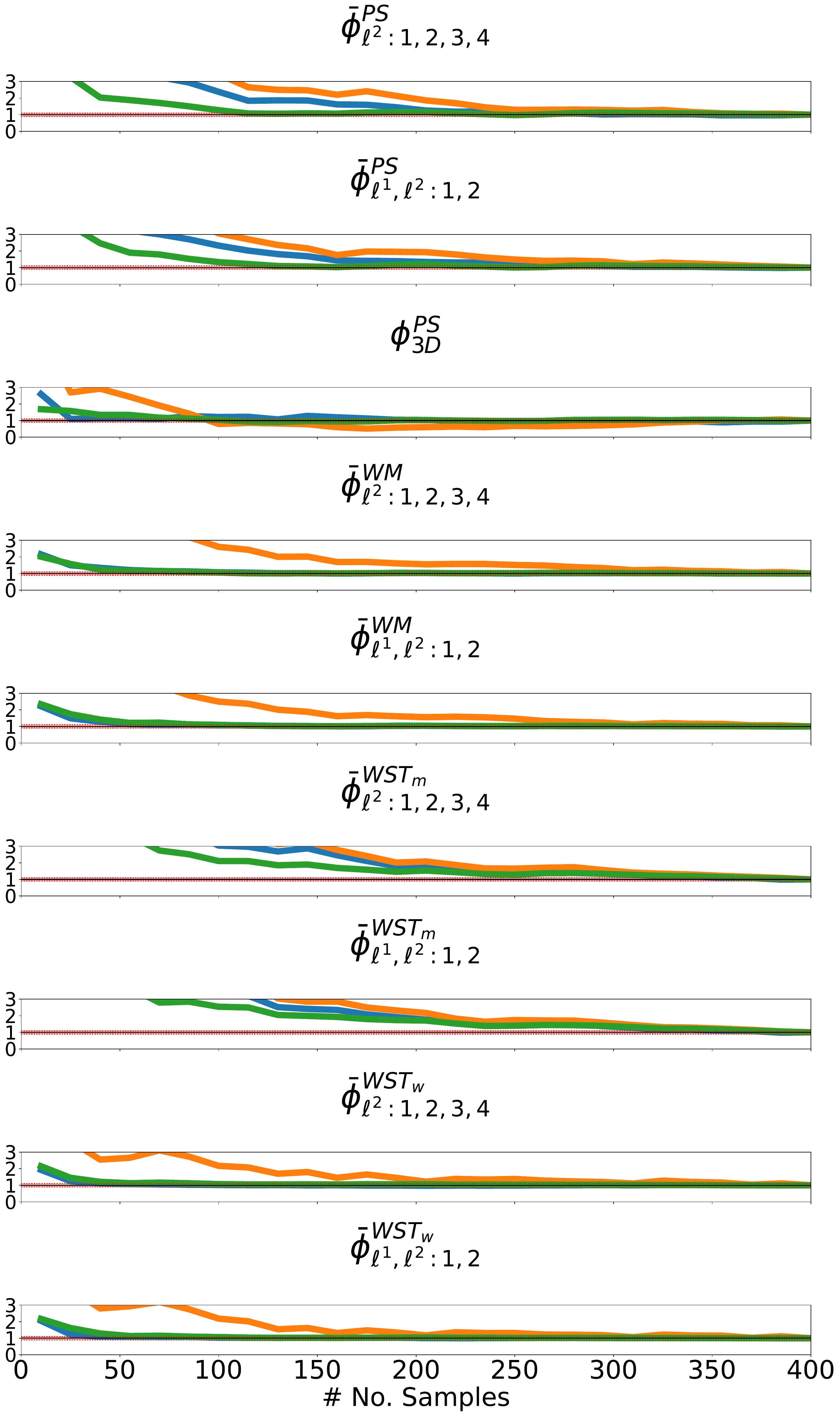}
 \caption{100 hours of SKA noise}
 \label{subfig:Full_Noiseless_convergence_100hrs}
\end{subfigure}%
\begin{subfigure}{.35\linewidth}
 \centering
 \includegraphics[width=\linewidth]{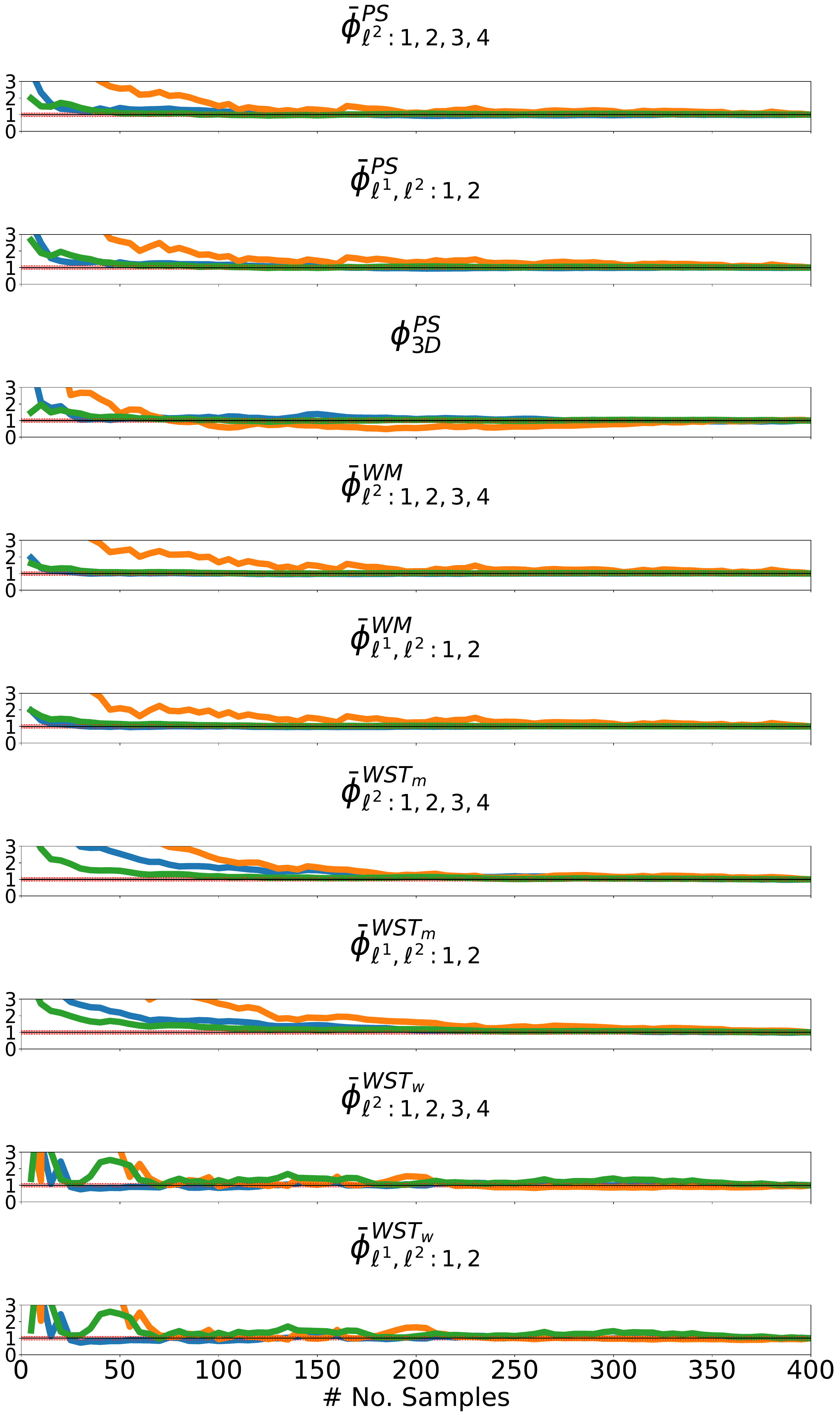}
 \caption{1000 hours of SKA noise}
 \label{subfig:Full_Noiseless_convergence_1000hrs}
\end{subfigure}
\caption{Convergence plots using equation \eqref{eq:Convergence}, where we have kept the number of fiducial simulations for our covariance constant at 600 and are solely varying the number of samples used for the derivatives. We consider the different noise cases: (a) no noise, (b) 100 hours of SKA noise, and (c) 1000 hours of SKA noise. We see that by 400 samples all of our statistics are fully convergent, falling within the 10$\%$, shown as the red shaded region. }\label{fig:Convergence_plots}
\end{figure*}

In this test, we keep the number of fiducial simulations used to calculate the covariance constant at 600, and vary the number of simulations used to calculate the derivatives from 1 to 400.

We can see from Fig.~\ref{subfig:Full_Noiseless_convergence} that all of our different statistics are fully converged by 400 samples, with $T_{Vir}$ and $\zeta$ being the most readily convergent parameters, converging after only 50 or so samples. However, $R_{Max}$ requires more samples before it converges towards the `target range' ---though it is at the edge of the convergence region for 200 or so samples.
In Fig.~\ref{subfig:Full_Noiseless_convergence_100hrs}, we show the convergence plots of our statistics in a high-noise scenario. We see that all the statistics reach convergence after 200 samples. This is likely due to the fact that this is a noise-dominated regime, and so after only 100 samples, for all statistics, the Fisher matrix is at the 10$\%$ error before needing 100 samples more to fully converge.
In Fig.~\ref{subfig:Full_Noiseless_convergence_1000hrs}, we show the convergence plots of our statistics for 1000 hours of SKA noise. We see a similar trend to that in Fig.~\ref{subfig:Full_Noiseless_convergence_100hrs}, where the regime is still noise dominated in nature, and around 200 samples are required to reach noise convergence.
\end{document}